\documentclass[preprint,floatfix
,aps,prd,superscriptaddress]{revtex4}

\usepackage[utf8]{inputenc}
\usepackage{graphicx}
\usepackage{caption}
\usepackage{tikz}
\usepackage{amssymb,amsmath,mathtools,comment}
\usepackage{mathrsfs}
\usepackage{xcolor}
\usepackage{breqn} 

\newcommand{\R}{\mathbf{r}} 
\newcommand{\U}{\mathbf{u}} 
\newcommand{\V}{\mathbf{v}} 
\newcommand{\W}{\mathbf{w}} 
\newcommand{\Fc}{\mathcal{F}}
\newcommand{\Dc}{\mathcal{D}}
\newcommand{\intj}{\int_{(j-1)L/n}^{jL/n}} 

\makeatletter
\let\cat@comma@active\@empty
\makeatother

\begin{document}

\title{Field-extension statistics of charged semiflexible polymers stretched with uniform electric fields}
\author{Ananya Mondal}
\email[]{am182@rice.edu}
\affiliation{ Center for Theoretical Biological Physics, Rice University, Houston, Texas 77005, USA}
\author{Greg Morrison}
\affiliation{ Center for Theoretical Biological Physics, Rice University, Houston, Texas 77005, USA}
\affiliation{Department of Physics, University of Houston, Houston, Texas 77204, USA}
\date{\today}

\begin{abstract}

Single-molecule force-extension experiments have enabled the manipulation of biomolecules in unprecedented detail. These studies have allowed quantitative measurements of the mechanical responses of biomolecules to applied forces explaining their roles in key biological functions. Electrophoretic stretching of charged polymers such as DNA in uniform electric fields is one such example, currently, used for sequencing purposes. Field-extension statistics of charged polymers differ from laser optical tweezer setups for force-extension experiments due to a non-uniform tension along the backbone of the chain, the effects of which remain poorly understood. In this paper, we modify an existing analytically tractable mean-field (MF) approach to account for the heterogeneity in tension for electric fields. This model has been shown to successfully predict the force-extension relations of inextensible polymers under uniform tension. Naively using this model for stretching of charged polymers such as DNA under electric fields results in local overstretching of the chain and gives inaccurate field-extension statistics. We improve this approach and account for the inhomogeneity in the tension by subdividing the chain into smaller segments while imposing the inextensibility of the chain. We find that the subdivided MF model shows better agreement with the simulations for the force-extension plots, and can predict the extension of the chain at any applied field value. We also show that using an isotropic mean-field model overestimates the longitudinal fluctuations both for tension forces as well as for fields. We correct the quantitative predictions for the longitudinal fluctuations in the mean extension by numerically differentiating the field-extension plots. We also find that the subdivided MF model can accurately predict the statistics of experimentally relevant quantities, such as transverse fluctuations, due to the analytical tractability of the model. These field-extension predictions may be further used to introduce confinement effects in the subdivided MF model and develop a comprehensive understanding of sequencing technologies. 
\end{abstract}

\maketitle

\section{Introduction}

\subsection{Stretching of biomolecules}

Single-molecule force-extension experiments \cite{bustamante2003ten, smith1992direct} have permitted the manipulation of biomolecules, and studying their responses to external fields has provided useful insights into the behavior of these biologically relevant molecules. These experiments are often conducted in a constant force ensemble \cite{brower2002mechanical,cui2000pulling,kulic2004dna} (e.g. laser optical tweezers), and the force-extension relations are well described with a wormlike chain model \cite{rubinstein2003polymer} under uniform tension \cite{ha1995mean, marko1995stretching}. DNA is a polyelectrolyte (PE), and in many contexts, it is more convenient to stretch charged bio-molecules by applying a constant electric field. Electrophoresis \cite{stellwagen2009electrophoresis} has been a popular separation technique for sorting DNA based on its molecular size. This technique has also been shown \cite{gurrieri1999trapping,long1996stretching} to be useful for observation purposes of long DNA polymers in agarose, by pinning the molecule and elongating it using large electric fields.  

More recently, technical advancements have made it possible to create nanochannels \cite{yeh2016stretching,mahshid2018transverse,roy2017electric,dorfman2013beyond} and nanopores \cite{belkin2015plasmonic,shi2017nanopore} through which polyelectrolytes tethered at one end can be elongated under electric fields. The ability of confined channels to suppress thermal fluctuations \cite{chen2017conformational,suma2018electric,morrison2020correlation} and observe linear DNA has made electrophoretic stretching an active area of research with growing applications in nanopore-DNA sequencing \cite{belkin2015plasmonic,heerema2016graphene} and consequent identification of epigenetic modifications \cite{cortini2016physics}. Electric field-driven stretching of DNA inside nanopores is currently used in low-cost long-read genome sequencing devices. One of the major advantages of using the nanopore sequencing technique compared to stretching charged polymers like DNA through biological pores is the prevention of interactions with the pore surface. In this sequencing method, one end of the DNA molecule is pinned with optical tweezers and the other end is subject to a constant electric field through a nanopore. Confinement effects \cite{morrison2009semiflexible, chen2016theory} can also give rise to complex statistics like back folding or hairpin formation, or diffusion of knots. This technique holds promise in making gene sequencing affordable for individual patient diagnostics but a problem lies in the precision of nucleotide sequence reading is still an open problem \cite{heng2005stretching, qiu2015intrinsic, qiu2021nanopores,lu2011origins, wanunu2012nanopores}. To improve the precision it is important to understand the physical properties of the charged polymer while it is stretched under a field through molecular simulations and a mathematical model. 

Stretching of single biomolecules with a constant tension is well-studied in laser optical tweezer experiments \cite{bustamante2003ten} but unlike these systems, stretching with an electric field generates a non-uniform tension profile along the backbone as seen in experiments \cite{gurrieri1999trapping, roy2017electric, marko1995stretching, wanunu2012nanopores}. Electric fields \cite{liedel2012beyond, liedel2013block} and shear flows \cite{majewski2015block} induced alignment in artificial block co-polymers are also being used to study the effects of local stress fields on different morphology and improve nanofabrication techniques. A common feature reported in electric field stretching experiments \cite{klotz2017dynamics,roy2017electric} is inhomogeneous tension created along the backbone of the polyelectrolyte. The tension gradient increases from zero at the free end to a maximum at the tethered end of the polymer. The inhomogeneity in the tension depends upon the charge distribution on the charged polymer but even for a uniformly charged polymer, the chain develops a non-uniform tension profile. The charge distribution, however, impacts the screening effects from the condensation of counter-ions \cite{manning1969limiting} that results in an effective charge distribution on the chain. The effective screening of DNA is also influenced by the viscous drag from the surrounding solvent molecules. This effect is called an electro-osmotic flow \cite{ghosal2019solid, luan2008electro}, which is in the opposite direction to the field stretching. Since the hydrodynamic drag is competing with the electric field stretching so the strength of this force determines how much the chain is linearly stretched and the strength is determined by the confinement radius of the nanochannels. 

To utilize the full potential of electrophoretic stretching and sequencing technology, it is important to shed light on the statistical properties of polymers under inhomogeneous tension.  In this paper, we study a simpler case of charged polymers stretched in an electric field. As a stepping stone to more higher-order interactions, we devise an analytically tractable solution for predicting field-extension plots for a single semiflexible filament pinned at one end and stretched with an electric field. The mean-field model used in this paper describes the semiflexible biomolecule using the persistence length of the chain that can be described as an effective electrostatic persistence length \cite{skolnick1977electrostatic} based on the screened charge density on the chain. Hydrodynamic and confinement effects however are future avenues for this project.

\subsection{Mathematical models}

Many biomolecules and their assemblies are often described using a wormlike chain (WLC) \cite{rubinstein2003polymer}, or the related discrete Kratky Porod Model \cite{kratky1949rontgenuntersuchung, doi1988theory}, which account for the inextensibility and intrinsic resistance to bending of these polymers. The competition between entropy (which prefers the formation of a random coil) and the energetic cost of bending (which prefers a straight conformation) leads to statistics in the WLC that differ significantly from flexible polymer models\cite{rubinstein2003polymer}. The persistence length $l_{p}$ is a measure of the energetic cost required to bend the polymer away from a straight configuration. Despite the WLC model being physically reasonable, it is often formidable to extract equilibrium or dynamical properties analytically \cite{hori2007stretching, samuel2002elasticity}. In order to develop analytically tractable techniques to study the statistics of a WLC, a number of useful approximate theories have been developed over the years. 
Theoretical studies \cite{marko1995stretching, ha1997semiflexible,bucci2020systematic,hori2007stretching, jacobson2017single, benetatos2004linear, bleha2019force, everaers2021single} have used equilibrium statistical mechanics to model experimental observable such as the mean end-to-end extension of WLCs when pulled under tension. For strongly stretched WLCs under large forces, the chain only slightly fluctuates in the direction transverse to the applied force. This behavior is approximated as the weakly bending rod (WBR) limit\cite{benetatos2010stretching, marko1995stretching, andersen2021stretching}. Using the WBR limit, Marko and Siggia (MS) \cite{marko1995stretching} have derived an interpolated force-extension equation from low to high values of a uniform tension force, $f$, that remains widely used. At low forces, the chain extends linearly, much like a Hookean spring $\langle R_{ee} \rangle /L \propto f$, where $\langle R_{ee} \rangle $ is the mean end-to-end extension of the chain in the direction of the applied force $f$). For higher forces, where WLCs such as DNA are stretched between $30-95 \%$ of its contour length, the difference between the mean end-to-end extension and the total contour length goes to zero as $1/f^{1/2}$ \cite{marko1995stretching}, which is also accounted for in the MS interpolation. The deviation of a WLC from a freely jointed chain's extension is accurately captured by the interpolation equation  in \cite{marko1995stretching}. 

MS \cite{marko1995stretching} and others \cite{andersen2021stretching, petrosyan2017improved, bouchiat1999estimating}, have recognized that it is a harder problem to solve the interpolation equation for inhomogeneous forces such as a constant electric field. A reasonable approximation to make meaningful predictions for force extension due to an electric field uses the WBR limit as most of the chain is strongly stretched and oriented in the direction of the applied field. From this, one obtains a scaling for the strongly-stretched regime, where the difference between the mean end-to-end extension and the total contour length also goes to zero as $1/\epsilon^{1/2}$, with $\epsilon$ being a dimensionless electric field \cite{marko1995stretching}. Due to the similarity in scaling with tension, MS argued that the electric field can be treated as an effective uniform tension in the WBR limit. The main challenge is to then predict the force-extension statistics in the intermediate region when thermal fluctuation is comparable to electrostatic stretching energy. In this intermediate region, it would be useful to have a simple interpolation (similar to MS) or an analytical expression that predicts the extension of the polymer. Although an exact field-extension curve can be derived, the calculation is quite laborious \cite{hori2007stretching, bleha2019force}, and involves an infinite series of Airy functions. One of the aims here is to find a simpler formula for experimentalists to use to analyze their data. 

In this paper, we use an existing analytically tractable mean-field theory \cite{ha1995mean, ha1997semiflexible, morrison2009semiflexible} that has successfully predicted the force-extension statistics of a semiflexible polymer stretched with uniform tension forces and improve on the model to predict a force-extension formula for electric fields. Lagowski et al \cite{lagowski1991stiff}, proposed a functional integral formalism that uses a mean-field approach to yield analytically tractable results for wormlike chains without using the WBR limit. Winkler et al \cite{winkler1994models} have demonstrated a procedure that calculates the partition function for discrete stiff chains exactly, using the maximum entropy principle. The mean-field model was further adopted by Ha and Thirumalai \cite{ha1995mean, ha1997semiflexible} and applied to the problem of semiflexible polymers stretched using uniform tension. The mathematical convenience of the mean-field method has proved fruitful in reproducing the scaling given by Marko and Siggia \cite{marko1995stretching} and also in creating a recipe to examine other useful quantities such as fluctuations and correlation functions. Furthermore, the mean-field method has been successful in producing tractable theories for biologically relevant systems such as kinetics of loop formation in WLCs \cite{hyeon2006kinetics} and confinement of biopolymers \cite{morrison2009semiflexible, morrison2020correlation}.  Benetatos and Frey \cite{benetatos2004linear} have made an interesting contribution to calculating the linear response of a grafted polyelectrolyte to a uniform electric field using spherical harmonics for different limiting conditions. Hori et al \cite{hori2007stretching} have studied the stretching of short biopolymers using electric fields in the WBR limit and have shown finite-size effects have a significant role to play in the field-extension statistics. Inspired by these theoretical studies, in this paper, our objective is to construct a minimally formidable mean-field technique that not only provides a pathway to derive field-extension statistics of WLCs in electric fields but also comes up with a workable formula for experimentalists to test our theory.

This paper begins with an introduction to a mean-field model \cite{ha1995mean,morrison2009semiflexible} used for predicting the force-extension statistics for worm-like chains stretched with uniform forces. In Section \ref{undivMF.sec} we describe the mathematical framework of the MF model in the context of WLCs stretched in electric fields, and we demonstrate that the MF model does not account for the inhomogeneity in the tension for the case of the electric field stretching. We show the underlying homogeneity assumption in the stretching parameter in the MF model leads to overstretching of the bonds and hence fails to replicate the experimental behavior of WLCs in a high electric field stretching. To improve the MF model we propose a `subdivided mean-field (MF) theory' in Section \ref{subdiv.sec}, dividing the chain into subdivisions corresponding to variable stretching parameters. The simulation methodology is described in Section \ref{simsel.sec}. In Section \ref{resultsel.sec} we predict the wide range of statistics, including field-extension plots, longitudinal fluctuations, and transverse fluctuations using the `subdivided MF model'. We find that simulations agree better with the field-extension statistics proposed here because in this model, the inhomogeneous tension is correctly accounted for. In Section \ref{longfluc.sec}, we show that the longitudinal fluctuations predicted from the MF approach are overestimated due to the isotropy in MF model. We also compare the FEC plots with the widely used interpolation formula proposed by Marko et al. \cite{marko1995stretching}, where the underlying assumption is an effective force can replace the electric field. In Section \ref{compare.sec}, we show that the assumption of the Marko-Siggia (MS) equation is not accounted for intermediate values of the electric field where the inhomogeneity in the tension is most pronounced. Finally, we suggest a new method to improve the prediction of longitudinal fluctuations in Section \ref{compare.sec}. We conclude with a summary of our results, applicability, and limitations.

\section{\label{methel.sec}Methods}

\subsection{\label{undivMF.sec}Homogeneous mean-field (MF) model for WLCs stretched in a uniform electric field}

\begin{figure}[htbp!]
	\begin{center}
		\includegraphics[width=0.6\linewidth]{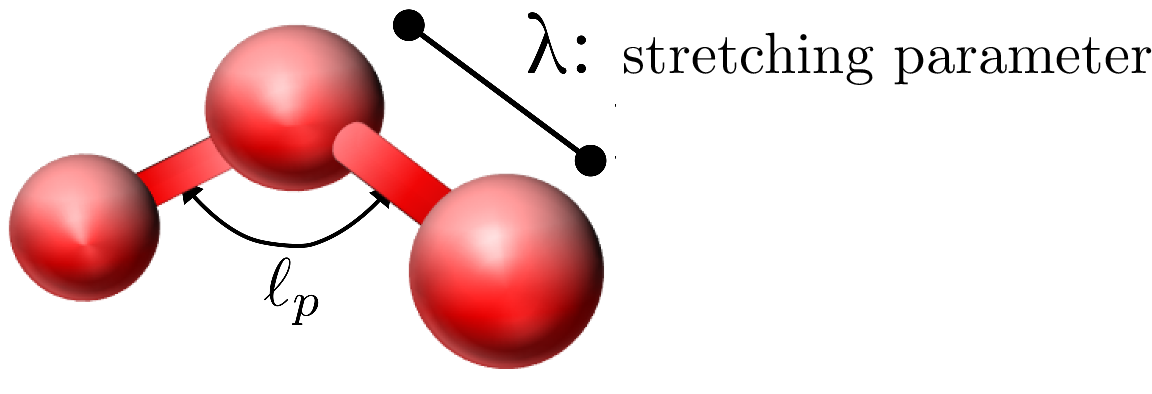}
		\caption[Coarse-grained bead spring model for worm-like chains (WLCs)]{\textbf{Coarse-grained bead spring model for worm-like chains (WLCs)} -- The coarse-grained bead spring model for worm-like chains has two energetic contributions, namely, bending energy represented by the persistence length ($l_p$) and stretching energy between the bonds represented by the parameter $\lambda$. }
		\label{mf.fig}
	\end{center}
\end{figure}

The Gaussian bead-spring model \cite{doi1988theory} is the simplest mathematical representation for long flexible polymers. Many biomolecules are semiflexible having an internal stiffness that needs to be accounted for in the model. The worm-like chain (WLC) model has been used to describe the statistics of biomolecules that are semiflexible like DNA, and actin. An inextensible wormlike chain, parameterized by the arc length $s$ and with position $\mathbf{r}(s)$ has the Hamiltonian $H_{wlc}=\frac{l_p}{2}\int_0^L ds (\partial_s\hat{\mathbf{\U}})^2$, 
with $\hat{\mathbf{\U}}=\partial_s\mathbf{\R}/|\partial_s\mathbf{\R}|$ the local unit tangent vector and $\partial_s$ a derivative with respect to $s$.  The WLC Hamiltonian represents the resistance to bending, with stiffer chains having longer persistence lengths $l_p$ (where $l_p=\kappa/k_B T$ where $\kappa$ is the bending modulus of the chain and $T$ the temperature).  Statistical averages using this Hamiltonian must account for the constraint of fixed length $L$, and numerous papers \cite{hori2007stretching, lagowski1991stiff} have determined the statistics of WLCs in various potentials (usually expressed as a series expansion that cannot easily be reduced to elementary functions).  In many cases, relaxing the rigid constraint of inextensibility permits analytically tractable predictions in terms of elementary functions, such as the Weakly Bending Rod (WBR) and Mean Field (MF) approaches.  In the MF approach, the rigid bond constraints are replaced by an average length and are introduced in the Hamiltonian as the inextensibility constraint using Lagrange multipliers. This assumption of averaging out the bond lengths is referred to as a Mean-Field (MF) model and it has successfully predicted experimentally relevant quantities in several studies \cite{ha1995mean, morrison2009semiflexible, winkler2003deformation, morrison2020correlation, mondal2022compression} involving biopolymers. 

For a charged semiflexible chain of uniform linear charge density $\sigma$, and length $L$, placed in an electric field $\boldsymbol{E}= E \boldsymbol{\hat x}$, the electrostatic energy is $-\sigma\beta \mathbf{E}\cdot\int_0^L ds \ \R(s) = -\sigma\beta \mathbf{E} \cdot \int_0^L ds \int_0^s ds' \U(s')$. In this section, we use the MF model that works well for uniform tension and use it na\"ively for the electric field. We find that an underlying homogeneity assumption in the stretching resistances of the bonds leads to the failure of the MF model in this case. The homogeneity assumption in the MF model causes the bonds near the tether to overstretch compared to the bonds near the free end because of a non-uniform tension profile due to the electric field. The stretching resistances do not account for inhomogeneous stretching and so we hypothesize the homogeneous MF model fails to capture the correct statistics in this case. We improve upon this model in section \ref{subdiv.sec} but before we do that let us take a closer look into the homogeneous MF model Hamiltonian. The MF Hamiltonian for a charged WLC in an electric field is 
\begin{eqnarray}
	\beta H_0[\U(s)]=\delta_0 \U_0^2+ \delta_L \U_L^2 +\frac{l_0}{2}\int_0^L ds \dot \U^2+\lambda\int_0^L ds \U^2 -  \boldsymbol{\epsilon} \cdot \int_0^L ds \int_0^s ds' \U(s')
 \label{mfham1}\;, 
\end{eqnarray}
where $l_0$ is a `mean-field persistence length' ($l_0\ne l_p$), $\lambda$ is the Lagrange multiplier to introduce the mean-field inextensibility constraint as resistance to stretching, $\U = \dot{\R}$ is the local stretching of the chain in the continuum limit and $\epsilon =  \sigma \beta E$ is the dimensionless electric field force. The chain is pinned at $\R(0)$ end to avoid a tumbling \cite{smith1999single} or formation of knots \cite{tang2011compression}. For a positively charged chain ($\sigma > 0$) the charges align in the direction of the electric field, stretching out the chain linearly. In the absence of an electric field, the terms involving $\delta$ are required to recover the known bending correlation function \cite{morrison2020correlation} using the MF approach, suppressing excess endpoint fluctuations. Pinning of one end causes an asymmetry in the fluctuations at each endpoint and so, the Lagrange multipliers or resistance to stretching parameters at the endpoints are respectively, $\lambda_0 = \lambda + \delta_0$ at the pinned end and $\lambda_L = \lambda + \delta_L$. The MF approach chooses the parameters $\lambda$, $\delta_0$ and $\delta_L$, such that the global inextensibility condition on the length is imposed on average, with $\langle \int_0^L ds \U^2 \rangle_0 = L$, $\langle \U^2_0 \rangle_0 = 1$, and $\langle \U^2_L \rangle_0 = 1$ (with $\langle ... \rangle_0$ is the statistical ensemble average using the MF Hamiltonian from Equation \ref{mfham1}). The mean-field persistence length $l_0$ for any freely fluctuating WLC and its relationship with the true persistence length $l_p$ can be determined using exactly known bond correlations \cite{ha1995mean, ha1997semiflexible, morrison2020correlation}. For three-dimensional, the relation is $l_0 = 3l_p/2$ but for two-dimension, the relation is more complicated. 

Unlike stretching by a uniform tension as is generally calculated in the literature \cite{ha1995mean}, stretching a WLC in uniform electric field results in a non-uniform tension profile as also reported in experiments \cite{gurrieri1999trapping, roy2017electric, stellwagen2009effect,keyser2006direct,maier2002elastic}. The inhomogeneity in tension along the backbone originates from the electrostatic energy term, $ U_{\epsilon} [\R] = -\boldsymbol{\epsilon}\cdot\int_0^L ds \ \R(s)$. For a positively charged chain, the dimensionless energy at an arc length of $s_1<s_2$ is $-\boldsymbol{\epsilon}\cdot\int_0^{s_1} ds \ \R(s) > -\boldsymbol{\epsilon}\cdot\int_0^{s_2} ds \ \R(s)$ is true on average, but not for all configurations like for a low electric field value. It is energetically more favorable for the chain to be aligned with the field on the tethered end as this ensures aligning a higher density of positive charges with the direction of the field. The electric field term can be further simplified by reversing the order of the double integrals and this results in the term shortening to $-\boldsymbol{\epsilon}\cdot \int_0^L ds (L-s) \U$, with $\U = \partial_s \R(s)$. The inhomogeneity in the tension is also visible from the $\sigma (L-s)$ term because again for $s_1 < s_2$, the total positive charges that are not aligned with the field are greater for $\sigma (L-s_1)$ than for $\sigma(L-s_2)$. This is equivalent to the same argument that there is a non-uniform tension profile along the backbone giving rise to a more aligned chain near the tether and the other end is comparatively more freely fluctuating. In contrast, for optical tweezers experiments, the pulling force applied is generally uniform, which necessarily means each point on the polymer chain feels the same amount of pull. 

Combining $U_{\epsilon} [\R]$ with Equation \ref{mfham1}, produces a quadratic Hamiltonian that is straightforward to work with. We complete the squares using the transformation, $\V(s) = (v_x,v_y,v_z)= ( u_x - \epsilon (L-s)/2\lambda,u_y,u_z)$ and $\dot{\V}(s) = (\dot{u_x} + \epsilon/2\lambda,\dot{u_y},\dot{u_z})$. The transformed MF Hamiltonian is given by
\begin{align}
	\beta H_0[\V(s)]= \delta_0 & \left(v_x(0)  + \frac{\epsilon L}{2\lambda}\right)^2  + \delta_0 \left(v_y(0)^2 + v_z(0)^2 \right)  + \delta_L \V_L^2 +\frac{l_0}{2}\int_0^L ds \dot \V^2+\lambda\int_0^L ds \V^2 \nonumber \\
 &- \frac{l_0\epsilon}{2\lambda}(v_x(L) - v_x(0)) - \int_0^L ds \ \frac{\epsilon^2(L-s)^2}{4\lambda} + \int_0^L ds \ \frac{l_0 \epsilon^2}{8\lambda^2}
 \label{mfham2}\;.
\end{align}

The quadratic Hamiltonian makes free energy calculation convenient for a WLC,  $\Fc_0=-\log\left[\int \Dc[\V(s)]e^{-\beta H_0}\right] - \lambda L - \delta_0 - \delta_L$, with the Lagrange multipliers given by free energy minimization conditions, $\partial \Fc/\partial \lambda = \partial \Fc/\partial \delta_0 = \partial \Fc/\partial \delta_L = 0$.
The free energy calculation reduces to the well-known classical path integral for a harmonic oscillator \cite{feynman2018statistical}. The free energy is a $d$-dimensional Gaussian integral of the form $-\log \left[\int \Dc [\V(s)]\  \exp(-\ \mathbf{V}^T\mathbf{A}\mathbf{V}) \exp( \mathbf{B}^T\cdot \mathbf{V} + \mathbf{C}) \right]$, where $\mathbf{A}$ is a tri-diagonal coefficient matrix of the quadratic terms, $\mathbf{B}$ is the coefficient matrix of the linear terms, $\mathbf{C}$ is a constant term matrix (a detailed path integral calculation in the Supplemental Information), and $\mathbf{V} = (\V_0,\V_L)^T$ is the column matrix for the endpoint variables. 

For zero electric field, the scaling is the same as any force-free WLC \cite{ha1995mean} and the mean-field conditions give, $\sqrt{\lambda l_0/2} = \delta = 3/4$. For the intermediate electric field values, the scaling is only determined numerically as this region has a mix of power laws. In the high electric field region, the scaling for the Lagrange multipliers with the electric field is obtained from the mean-field conditions, $\lambda \sim L\epsilon/2\sqrt{3}$, $\delta_0 \sim 0.341 \sqrt{Ll_p\epsilon}$, and $\delta_L \sim -0.465 \sqrt{Ll_p\epsilon}$, where $l_p$ is the true persistence length of the chain. The scaling of the Lagrange multipliers with the electric field is obtained from the numerical solutions of the mean-field equations. The difference in the constraining conditions for the two ends of the chain is reflected in the Lagrange multipliers $\delta_0$ and $\delta_L$, which, for sufficiently high fields, is either positive or negative.  A positive Lagrange multiplier or stretching parameter imparts harder resistance to stretch compared to a negative value.

\subsection{\label{overstretch.sec}Overstretching due to homogeneity assumption in the MF model }
 We expect that the homogeneous MF model does not account for the inhomogeneity in tension for WLCs stretched in an electric field and leads to over-stretching and compression of the bonds. It is energetically favorable for positively charged monomers to move in the direction of the field, causing the bonds to align with the electric field. While aligning with the electric field, the bonds do not stretch equally as they would have done if the polymer was pulled using a uniform tension \cite{marko1995stretching, ha1997semiflexible}. For a positively and uniformly charged polymer stretched in an electric field directed away from the tethered end as shown in Figure \ref{subdivided.fig}, the bonds closer to the tethered end stretch more compared to the bonds closer to the free end. The bonds near the tether prefer to stretch and not compress because, in the latter scenario, a larger number of the positive charges will be pulled out of the electric field making it an energetically unfavorable configuration. We define overstretching of the bonds as the extension of the local bond lengths beyond the fixed lengths while maintaining the global inextensibility of the chain. The overstretching effect of the electric field gives rise to an inhomogeneous stretching or tension-pulling profile. This inhomogeneous tension along the backbone of the chain is not a characteristic of the tether but is retained even in the bulk of long chains. The MF model assumes that the stretching parameters are equal for all bonds irrespective of their location in the electric field meaning that the bonds closer to the tether experience greater stretching forces locally, and can over-stretch for intermediate or high values of electric field. Similarly, the bonds closer to the fluctuating end experience less stretch and hence are effectively compressed.
 
\begin{figure}[htb!]
	\begin{center}
		\includegraphics[width=0.8\linewidth]{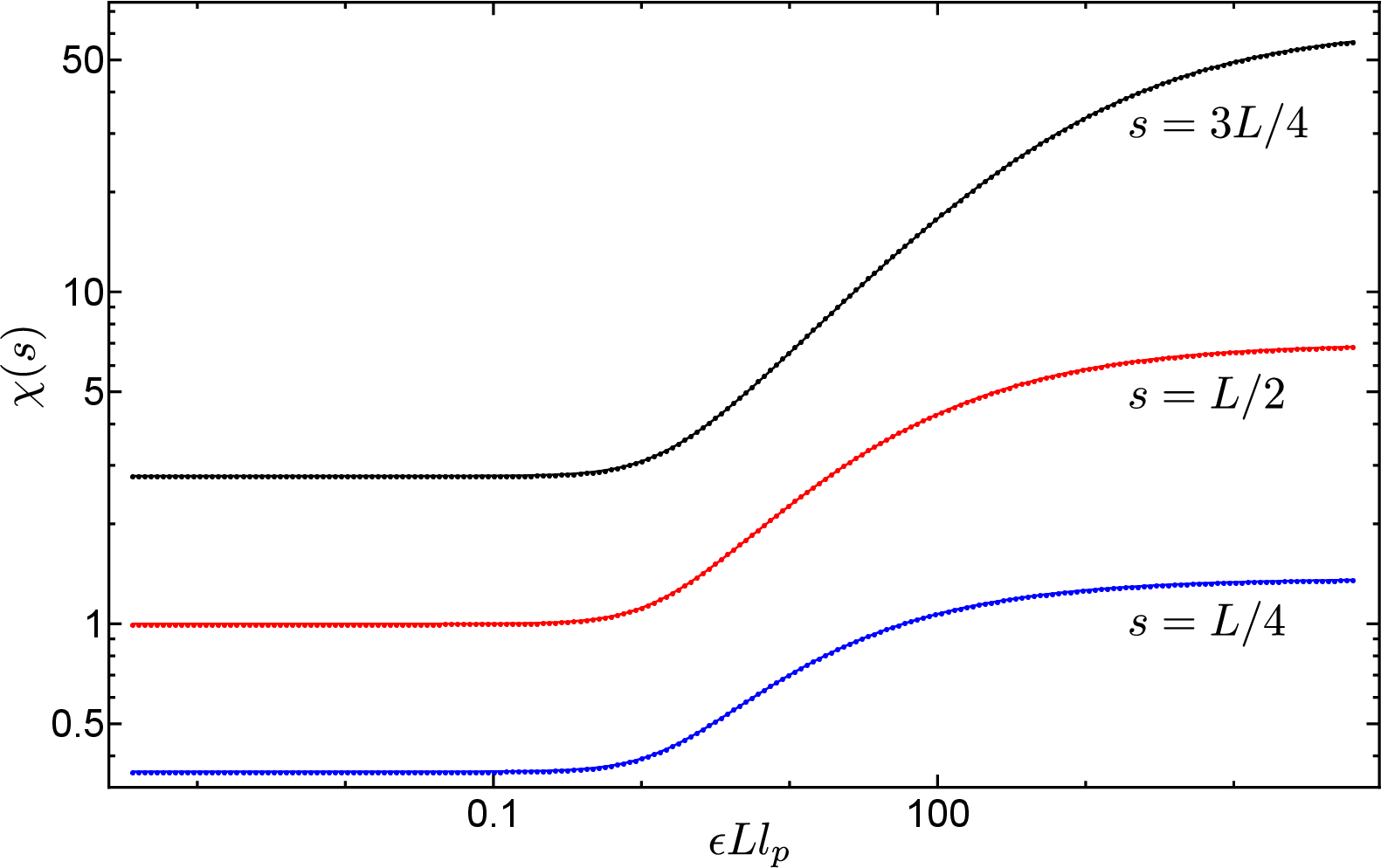}
		\caption[Homogeneity assumption in the mean-field model leads to overstretching of bonds]{\textbf{Homogeneity assumption in the mean-field model leads to overstretching of bonds} -- Bond correlation ratio, $\chi(s) = \langle \int_0^s ds^{'}  u_x^2(s') \rangle/\langle \int_s^L ds^{'}  u_x^2(s') \rangle$ plotted against the dimensionless field $\epsilon L l_p$ for 3 different values of the arc lengths, s = L/4 (blue), s = L/2 (red), s = 3L/4 (black). Points indicate numerical solutions for the quantity $\chi(s)$ obtained from the $n = 0$ mean-field model in Equation \ref{mfham1}. Overstretching of the bonds is a result of the mean-field assumption of homogeneous stretching parameters, $\lambda(s) = \lambda$, in the case of electric fields. For all 3 cases, the ratio picks up as the field is increased, which indicates that the bonds closer to the tethered end (s = 0) are stretched more compared to the other half of the chain closer to the free end (s= L). The plots show that the ratio of the overstretching and compression of the bonds is greater than 1 as $s$ is increased from $L/4$ to $3L/4$. This means that the overstretching is not compensated by the compression of the bonds or in other words, the chain does not attain full extension even for high electric fields.  }
		\label{over.fig}
	\end{center}
\end{figure}

To quantify the overstretching or compression in bonds as a function of the applied electric field, we evaluate the bond correlation ratio, $\chi(s) = \langle \int_0^s ds^{'}  u_x^2(s') \rangle/\langle \int_s^L ds^{'}  u_x^2(s') \rangle$ for different values of the arc length $s$ on the chain. The quantity, $\chi(s)$, is also calculated using the auxiliary field method (detailed in section \ref{extension.sec}), where $-\alpha_1  \int_0^s ds^{'}  u_x^2(s')$ and $-\alpha_2\int_s^L ds^{'}  u_x^2(s')$ terms are introduced to the undivided chain Hamiltonian in Equation \ref{mfham1} as before. Taking a derivative of the free energy with respect to the corresponding auxiliary field, $\partial (-\Fc^{\alpha})/\partial \alpha_{i} \Big{|}_{\alpha_{i} = 0}$, yields the ratio. This ratio is plotted for different values of $s$ in Figure \ref{over.fig}. The ratio of all the curves for different arc lengths is initially low for low fields and then picks up as the field is increased and finally saturates because of the contour length inextensibility constraint. A low ratio means that the bonds are equally stretched as expected for low fields. For higher values of the ratio, the bonds nearer to the tethered end are stretched more compared to the ones beyond the arc length $s$ or closer to the free end at $s = L$. The plots are shown for three different values of $ s = L/4, L/2, 3L/4$ to observe the effect of overstretching and compression due to the underlying homogeneity assumption in the undivided MF model. {\color{black} For $s = L/4$, $\chi < 1$ for $\epsilon \rightarrow 0$ and reaches $\chi \approx 1$ for $\epsilon \rightarrow \infty$ indicating overstretching and compression of the bonds on either side of $s = L/4$ of a polymer in an applied field. The $s = L/2$ (red) and $s = 3L/4$ (black) plots show that the overstretching and compression ratio of the bonds is much greater than one for high electric field values and even for low and intermediate ranges of the field. This shows that the homogeneous MF model without subdivisions cannot give the correct statistics for the field-extension curves. }

\subsection{\label{subdiv.sec}Subdivided chain mean-field model for WLCs stretched in a uniform electric field}

The end-to-end extension for WLC is often used as an observable in force-extension experiment \cite{bustamante2003ten} and plotted with the stretching force to determine the mechanical properties of WLCs. For a charged WLC stretched in an electric field, our aim is also to quantitatively predict the mean-extension and fluctuation statistics of the WLC as a function of the applied field. The MF model homogeneity assumption in the stretching parameter, $\lambda(s) = \lambda$, causes all the bonds to resist stretching equally, which means it fails to balance the inhomogeneity in the backbone tension due to the electric field. Bonds near the pinned end overstretch and bonds closer to the free end compress for large electric fields with $\lambda$ as a constant stretching resistance. The homogeneity assumption works well for uniform tension \cite{ha1995mean} but fails to account for the inhomogeneity in tension for electric fields. The inhomogeneity is not exactly accounted for in the homogeneous MF model, and the chain does not stretch to its full extension, $L$, even for a high field because the over-extension does not compensate for the compression for non-uniform stretching forces.

\begin{figure}[htbp!]
	\begin{center}
		\includegraphics[width=0.6\linewidth]{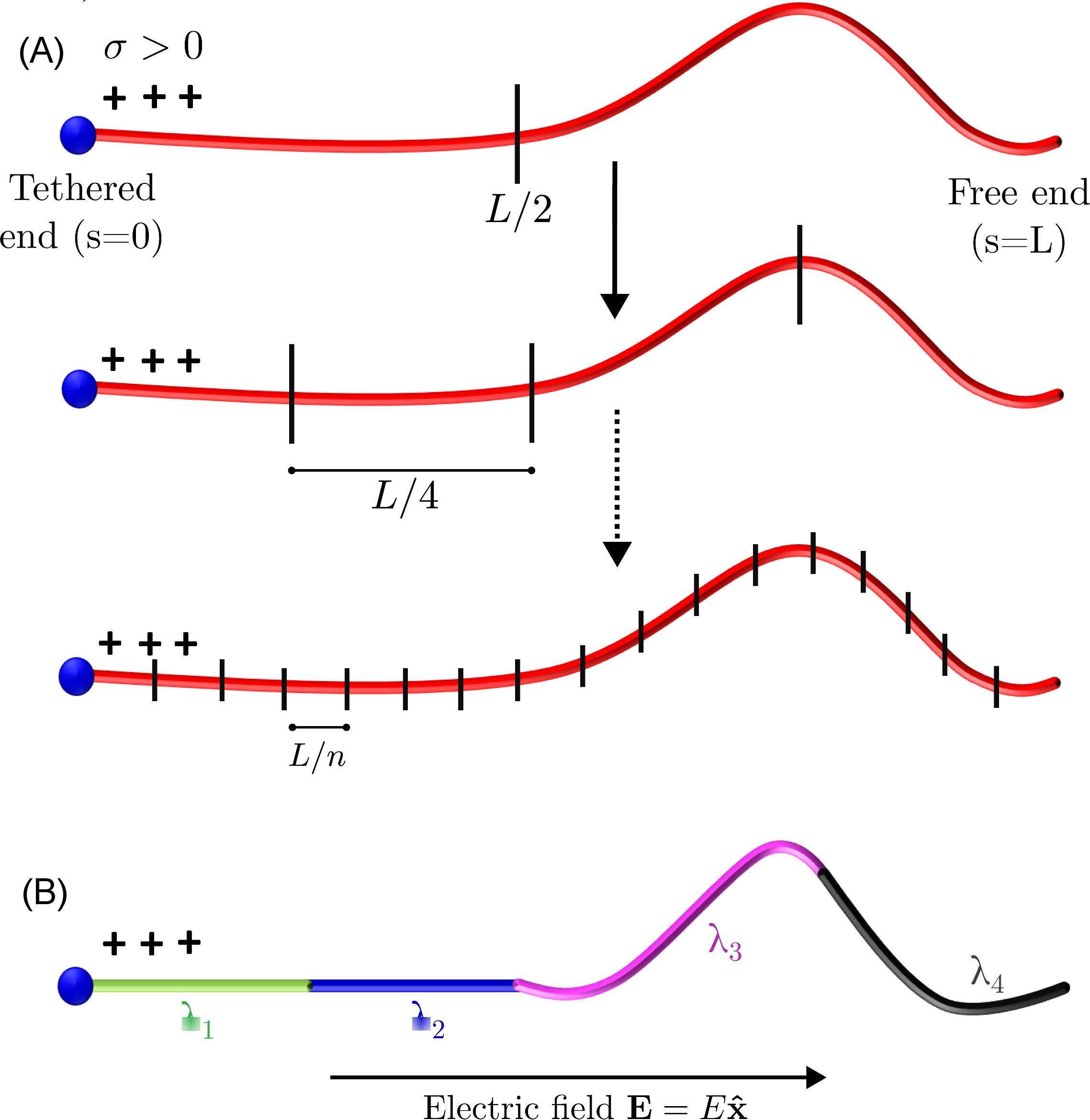}
		\caption[Schematic diagram of the subdivided mean-field (MF) theoretical model accounting for the inhomogeneity in the tension due to electric field stretching]{\textbf{Schematic diagram of the subdivided mean-field (MF) theoretical model accounting for the inhomogeneity in the tension due to electric field stretching} -- (A) A positively charged worm-like chain of length $L$ is tethered at one end (blue bead) and is freely fluctuating on the other. A uniform electric field, $\mathbf{E} = E \boldsymbol{\hat{x}}$ stretches the chain as the positive charges align in the direction of the field. The chain is stretched more on the tethered end than the free end because of inhomogeneity in tension. To account for the inhomogeneity in tension, we divide the chain into subdivisions, where each subdivision has a different stretching resistance. The length of each subdivision becomes vanishingly small for a large number of subdivisions. (B) The variable stretching parameters for 4 subdivisions are shown in $n = 4$ different colors, where $\lambda_1 > \lambda_2 > \lambda_3 > \lambda_4$ for the $n = 4$ subdivisions. The increasing order of the parameters is determined as a result of stationary phase conditions.}
		\label{subdivided.fig}
	\end{center}
\end{figure}

To improve on the homogeneous MF model, we propose a subdivided MF model as shown in Figure \ref{subdivided.fig}. The length of each subdivision is $L/n$, where $n$ is the total subdivisions. Dividing the chain means that the MF assumption on the bonds is now locally applied over each subdivision. Physically, this leads to an over- or under-stretch of the subdivisions but by making the regions smaller and smaller the effect of over or under-stretching is expected to be mitigated. Every region now over- or under-stretches, but the net effect is that the inextensibility is globally satisfied (even if not locally). The mean field assumption for the $j^{th}$ subdivision is given by $\langle \int_j ds_j \U_j^2 \rangle = L/n$, where $\int_j \equiv \intj$ and $\U_j$ is the stretching in the $j$th subdivision. The subdivided chain MF Hamiltonian from Equation \ref{mfham1} becomes
\begin{eqnarray}
	\beta H_0[\U_j(s)]=\delta_0 \U_0^2+ \delta_L \U_L^2 +\frac{l_0}{2}\int_0^L ds \dot \U^2 + \sum_{j=1}^n \lambda_j\int_j ds_j \U_j^2 -  \boldsymbol{\epsilon} \cdot \int_0^L ds \int_0^s ds' \U(s')
 \label{mfsub1}\;, 
\end{eqnarray}
where $\int_j$ is the integral over the segment $s = (j-1)L/n$ to $s = jL/n$ and $\lambda_j$ is the stretching parameters for the $j^{th}$ subdivision.

Completing the squares on the Hamiltonian in Equation \ref{mfsub1}, with the following transformation for the subdivisions, $\V_j = (v_{xj}, v_{yj}, v_{zj}) = (u_{xj} - (L-s)\epsilon/2\lambda_j, u_{yj}, u_{zj})$, results in the new Hamiltonian

\begin{align}
	\beta H_0[\V_j(s)] = & \sum_{j=1}^n \left(\frac{l_0}{2}\int_j ds_j \dot \V_j^2 + \lambda_j\int_j ds_j \V_j^2 - \int_j ds_j \ \frac{\epsilon^2(L-s_j)^2}{4\lambda_j} + \int_j ds_j \ \frac{l_0 \epsilon^2}{8\lambda_j^2} \right) \nonumber \\
 & -\sum_{j=1}^n \left( \frac{l_0\epsilon}{2\lambda_j}v_x(s = jL/n) + v_x(s = (j-1) L/n) \right) \nonumber \\
 & + \delta_0 \left(v_x(0) + \frac{\epsilon L}{2\lambda_1}\right)^2 + \delta_0 \left(v_y(0)^2 + v_z(0)^2 \right) + \delta_L \V_L^2
 \label{mfsub2}\;.
\end{align}
The free energy of the subdivided MF Hamiltonian in Equation \ref{mfsub2} is calculated using Gaussian integrals, $\Fc_{n}=-\log\left[\int \Dc[\V(s)]e^{-\beta H_0[\V_j(s)]}\right] - \sum_{j=1}^n \lambda_{j} L/n - \delta_0 - \delta_L$. The path integral calculation is detailed in Supplemental Information. The distribution of the Lagrange multipliers follows from the mean-field equations, $\partial \Fc_n/\partial \lambda_j = \partial \Fc_n/\partial \delta_0 =  \partial \Fc_n/\partial \delta_L = 0 $. 
 From the mean-field equations we obtain that in the zero electric field limit, all the stretching parameters are equal, $\lambda_j = 9/8l_0, \delta_0 = 3/4, \delta_L = 3/4$, which is expected for a WLC \cite{ha1995mean}. 
 \begin{figure}[htb!]
	\begin{center}
		\includegraphics[width=0.85\linewidth]{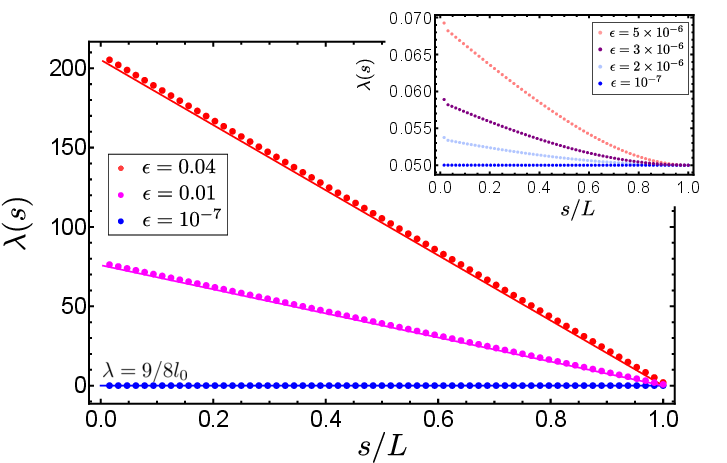}
		\caption[Variation of stretching parameters derived from the subdivided mean-field model as a function of position on the WLC]{\textbf{Variation of stretching parameters derived from the subdivided mean-field model as a function of position on the WLC} --Numerical solutions (solid dots) for $\lambda(s)$ as a function of arc length $s$ is plotted for different values of the fields. The lines are fit using the equation $\lambda(s) \sim (L-s)\epsilon/2$. For very low field (blue) the $\lambda$ values are constant for all subdivisions and are close to the zero field value of $\lambda_0 = 9/8l_0 = 0.05$ for $l_p = 15$. As the field is increased the variation in the stretching parameters for $n = 64$ subdivisions is observed. The variation is nonlinear for intermediate values of the field (shown in the inset). For high values of the field, the $\lambda(s)$ has a clear linear dependence on the arc length, and the stretching parameters closer to $s=0$ is higher compared to the ones closer to $s= L$. This order of the stretching parameters is a result of the inhomogeneous stretching tension on either side of the chain. A higher value of stretching resistance ensures that the bonds are harder to overstretch and a lower value of $\lambda(s)$ leads to less compression leading to correct extension statistics.}
		\label{lambda.fig}
	\end{center}
\end{figure}

 Figure \ref{lambda.fig} shows the variation of the stretching parameters for $n = 64$ subdivisions and different field values. The points are the numerical solutions, and solid lines are fits. For a low field, all the roots are the same and have a value very close to the values for zero fields, $\lambda = 9/8l_0$. For intermediate field values, the inset plots show the variation of the stretching parameters more clearly, as in this range of the field, the scaling laws are not very clear. With an increase in the field, the stretching parameters closer to $s = 0$, or the pinned end, are higher than for the subdivisions closer to the free end. It means that the bonds towards the pinned end are harder to stretch, and the variable distribution of $\lambda_j$ corrects the issue of over-stretching and compression of bonds. For a high value of the field, the trend of decreasing $\lambda_j$ remains, and it fits the scaling law $\lambda(s) \sim (L-s)\epsilon/2$ perfectly. We hypothesize that the issue of over-extension and compression of the bonds is corrected by using this distribution of stretching parameters, which makes it harder to overstretch bonds on the pinned end of the chain or compress bonds closer to the free end of the chain. We predict that accounting for the inhomogeneity in tension using the subdivided MF method results in full extension of the WLC in high electric fields.

\subsection{\label{uv.sec} Are the partition functions equivalent before and after the completion of squares? }

To obtain a quadratic Hamiltonian we complete squares in Equations \ref{mfham2} and \ref{mfsub2}. The partition functions calculated from the Hamiltonian before and after the completion of squares should be equal. To demonstrate the equivalence of the partition function before and after the completion of squares, the propagators in the partition functions (see Supplemental Information for details on propagators) must be equal as well. For example, in the subdivided chain, let us consider the propagators connecting the $j^{th}$ and $(j+1)^{th}$ subdivisions. For the adjacent propagators of $j^{th}$ and $(j+1)^{th}$ segments to be connected at the joining point, $v_{xj} = u_{xj} - (L\epsilon /2\lambda_{j} ) (1-j/n)$ from the $j^{th}$ segment and $v_{xj} = u_{xj} - (L\epsilon /2\lambda_{j+1})(1-j/n)$ from the $(j+1)^{th}$ segment, should be the same. The only difference between the connection point is the $\lambda_j$ or $\lambda_{j+1}$ parameter depending on if the $v_{xj}$ point is in the $j^{th}$ segment propagator or $(j+1)^{th}$ segment propagator. The connecting point, $v_{xj}$ in the $(j+1)^{th}$ propagator can also be rewritten as, $v_{xj} = u_{xj} - (L\epsilon /2)(1-j/n)(\frac{1}{\lambda_{j}}+\eta_j)$, where $\eta_j = (1/\lambda_{j+1} - 1/\lambda_j)$ is the difference in the inverse of the stretching parameters for the adjacent subdivisions. For a large number of subdivisions ($n \rightarrow \infty$), the difference $\eta_j \approx 0$ as the stretching parameters scale as $1/\lambda_j \sim 1/n$. It follows that for a large number of subdivisions, the connecting point $v_{xj}$ is the same for adjacent propagators and hence the equivalence of the partition functions holds true before and after the completion of squares. This equivalence is trivial for the non-subdivided chain.

\section{\label{simsel.sec}Simulations}
In order to quantify the predictive power of the subdivided chain model, we performed MCMC simulations to measure the mean and fluctuations of various observables. We simulate a coarse-grained chain of $N = 250$ monomers and the length of the chain is given by $L = (N-1)a$, where $a$ is the distance between monomers. The chain is generated by growing $(N-1)$ bond vectors, $\hat{\U_i}$, each with a fixed length. There are two energetic contributions: the bending energy from the intrinsic stiffness (persistence length $l_p$) of a worm-like chain and from the electric field. In our simulations, the bending energy is $U_{\text{bend}} = \kappa \sum_i (1 - \hat{\U}_i \cdot \hat{\U}_{i+1})$ with $\kappa$ as the bending stiffness parameter related to the persistence length, $l_p/a \approx (\kappa - 1 + \kappa \coth \kappa)/2(\kappa + 1 - \kappa \coth \kappa)$ \cite{dorfman2013beyond}, where $a$ is the distance between the beads. We used three different values of the persistence length in the simulations, $l_p =  2, 15, 30$ monomers. In our simulations, we take a positively and homogeneously charged chain with a charge density $\sigma$. The electric field energy is introduced in our simulations as a dimensionless field, $\epsilon = \sigma \beta E $, and varies from low values of $\epsilon = 10^{-5}$ to moderate and high values of $\epsilon = 1$, where the polymer stretches to its full contour length. The electric field energy in the simulation is $\Delta U_{E} = -\epsilon (N-i)$dx, $i$ is a randomly picked bond to be displaced to generate a trial configuration in the Monte Carlo simulations and dx is the change in the x-component of the bond for the trial configuration which is being aligned with the field.

Initially, each chain is grown randomly (without regard for the electric field).  At each Monte Carlo step a random bond $i$ is chosen uniformly along the chain, and replaced with a new trial at an angle $\theta_i$ with respect to the existing $(i-1)^{th}$ bond.  $\theta_i$ is drawn from $P(\theta_i)\propto e^{-a\cos(\theta_i)/l_p}$ (with no dependence on the electric field), and is accepted or rejected with the Metropolis criterion $p_{acc}=\mbox{max}(1,e^{-\beta \Delta U})$, with $U=U_{bend}+ \Delta U_{E}$. Using this algorithm we perform approximately $10^8$ MC steps between the initially grown chain and the final configuration. After equilibration, a total of $10000$ chains are generated to obtain field-extension statistics such as end-to-end distance or fluctuations at each combination of stiffness and applied field parameters.

\section{\label{resultsel.sec}Results}

\subsection{\label{extension.sec}Field-extension (FEC) plots}

Force-extension experiments measure the end-to-end extension of single molecules \cite{bustamante2003ten}, and can help determine the mechanical properties of biomolecules like persistence length. We calculate the mean end-to-end extension of the chain aligned with the direction of the field for low, intermediate, and high values of the applied electric field. We expect that the undivided MF model in Section \ref{undivMF.sec} does not predict the correct field-extension statistics due to overstretching/compression due to the inhomogeneous tension. The mean end-to-end extension in the direction of the field ($\mathbf{E} = E \boldsymbol{\hat{x}}$) is the sum of all the bond vectors aligned with the field, $\langle X_{ee} \rangle = \left\langle  \int_0^L ds u_x(s)  \right\rangle $. We calculate this quantity by adding an auxiliary field term, $-\alpha  \int_0^L ds u_x(s)  $ (see details in Supplemental Information), to the Hamiltonian in Equation \ref{mfham1}. Completing the squares for the new Hamiltonian using the transformation, $w_{x} = u_{x} - ((L-s)\epsilon + \alpha)/2\lambda$, enables us to calculate the average end-to-end extension analytically, $\langle X_{ee} \rangle = -\frac{\partial \Fc^{\alpha}}{\partial \alpha} \Big{|}_{\alpha = 0}$. For a long chain, the bulk of the chain contributes largely compared to the endpoints to the mean end-to-end extension, and in this limit, $\langle X_{ee} \rangle = \frac{L^2 \epsilon}{4\lambda}$. The stretching parameter, $\lambda$, is determined from the mean-field equation, $\frac{\partial \Fc^{\alpha}}{\partial \lambda} \Big{|}_{\alpha = 0} = 0$ and we neglect $\delta_0$ and $\delta_L$ as endpoint effects should not dominate the global extension and we neglect $\delta_0$ and $\delta_L$ as endpoint effects should not dominate the global extension. Simultaneously solving the mean-field equation for the stretching parameter, $\lambda$ and the mean end-to-end extension expression in the long chain limit, gives the field-extension equation for the undivided chain to order $1/L$,
\begin{eqnarray}
    -L + \frac{4 X_{ee}^2}{3 L}+3\sqrt{\frac{X_{ee}}{2 l_0\epsilon}} = 0.
    \label{exteq0}
\end{eqnarray}

Equation \ref{exteq0} has four solutions: two imaginary roots and one real root that results in a large value of $\langle X_{ee} \rangle$ exceeding the contour length $L$ (so these roots are excluded), and one real root that converges as $\epsilon \rightarrow 0$. In the high electric field limit, the mean end-to-end extension is $\langle X_{ee} \rangle = \sqrt{\frac{3}{4}} L$ for a $\lambda \sim L\epsilon/2\sqrt{3}$. The end-to-end extension is shorter than the full contour length for high electric fields due to local compression and overextension on the mean-field level. This is consistent with the result in Section \ref{overstretch.sec}.

The mean end-to-end extension for the $n$ subdivisions can also be calculated by introducing $n$ auxiliary field terms $-\alpha_j \int_j ds_j u_x(s_j)$ in the subdivided MF Hamiltonian with total extension, $\langle X_{ee} \rangle = -\sum_{j=1}^n \frac{\partial \Fc_n^{\alpha}}{\partial \alpha_j} \Big{|}_{\alpha_j = 0}$. From the calculation outlined in Supplemental Information, the free energy can be calculated and the mean end-to-end extension equation for the $j^{th}$ subdivision is obtained as $\langle X_{ee}^j \rangle = \frac{L^2\epsilon(1-2j+2n)}{4n^2\lambda_j}$. In the $L \rightarrow \infty$ limit the mean-field equations for the stretching parameters to order $1/L$ are, $-\frac{L}{n} + \int_j ds_j \ \frac{(L-s_j)^2\epsilon^2}{4\lambda_j^2} + \frac{3L}{2n\sqrt{2l_0\lambda_j}} = 0$. To determine the field-extension equations, we can readily solve for $\lambda_j = \frac{L^2 \epsilon(1-2j + 2n)}{4n^2\langle X_{ee}^j \rangle}$ and this results in the field-extension equations for the $j^{th}$ subdivision,
\begin{eqnarray}
    -\frac{L}{n} + \frac{4n(1+3j^2 + 3n(1+n) - 3j(1+2n)) (X_{ee}^j)^2}{3L(1-2j+2n)^2}  + 3\sqrt{\frac{X_{ee}^j}{(2-4j+4n)l_0 \epsilon}} = 0.
    \label{exteqn}
\end{eqnarray}
Exact expressions for $X_{ee}^j$ are unwieldy, but can be readily evaluated numerically.

{\color{black}In the limit of $L/n \rightarrow 0$, the total end-to-end extension of the chain becomes $\langle X_{ee} \rangle  = \int_0^L ds\  \frac{(L-s)\epsilon}{2\lambda(s)}$. The high field behavior of full extension is recovered with $\lambda(s) \sim (L-s)\epsilon/2$ leading to a full extension $\langle X_{ee} \rangle/L = 1$. Note that the expression, $\langle X_{ee} \rangle/ L = \frac{L\epsilon}{4\lambda}$, is very similar to the uniform stretching force extension derived by the authors in Ref. \cite{ha1995mean}, where $\langle X_{ee} \rangle/ L = \frac{fL}{2\lambda}$ except the coefficient of $\frac{1}{4}$ instead of $\frac{1}{2}$. The difference in coefficients is because of the inhomogeneous tension term in the electric field, $(L-s)$, as compared to the uniform stretching tension force $f$. The behavior of $\langle X_{ee} \rangle$ with respect to the field strength must be evaluated numerically as discussed in Section \ref{subdiv.sec}.}

\begin{figure}[htb!]
	\begin{center}
		\includegraphics[width=0.9\linewidth]{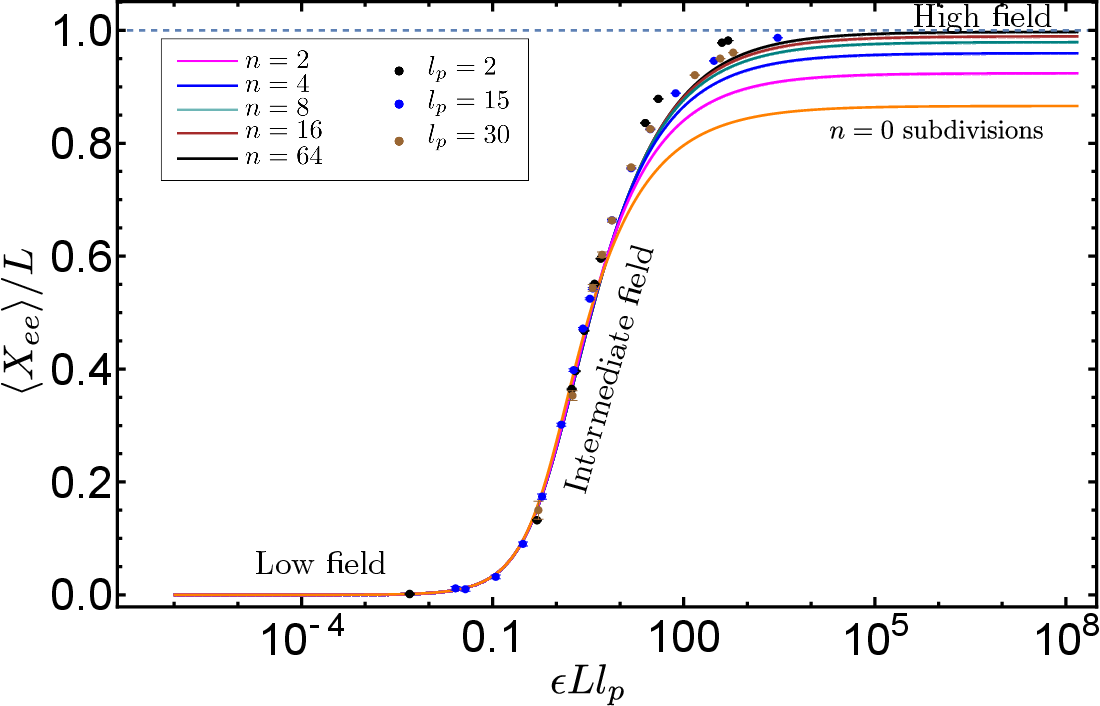}
		\caption[Field-extension curves for WLCs]{\textbf{Field-extension curves for WLCs} -- Normalized mean end-to-end extension $\langle X_{ee} \rangle/L$ of a worm-like chain stretched in a uniform electric field in the x-direction, is plotted as a function of the dimensionless electric field $\epsilon L l_p$, where $\epsilon = \sigma \beta E$. For low fields, the extension grows linearly but for intermediate fields, the extension is nonlinear due to the bending cost of worm-like chains. At very high values of the field, the chain stretches to its full contour length and the extension should saturate at 1 (dashed blue line). For $n = 0$ subdivisions (the homogeneous MF model without subdivisions), the high field limit is wrong as the chain only stretches to $0.86 L$. The saturation value improves as the number of subdivisions is increased. Simulation data shows good agreement with the numerical solution for the $n = 64$ curve (black solid line) for intermediate fields (even for small $n$). Simulations are shown for three different values of the persistence length, $l_p = 2 $ (black dots), $15 $ (blue dots), $30 $ (brown dots) monomers, and $L = 250$ monomers. Error bars are on the order of the symbol sizes.  }
		\label{fec.fig}
	\end{center}
\end{figure}

Figure \ref{fec.fig} shows the field-extension curves for different numbers of subdivisions plotted against the dimensionless electric field, $\epsilon L l_p$. As calculated using the MF approach for an undivided chain, the full end-to-end extension in the direction of the field is less than the total contour length, for high electric fields. Subdivision of the chain that accounts for the inhomogeneity in the tension for electric fields, improves the scaling of the extension for high fields. For $n = 64$ the total extension for high fields, $\langle X_{ee}/L \rangle \approx 1$, and in principle, for an infinite number of subdivisions the chain attains full extension as expected.

\subsection{\label{compare.sec}A comparative study with literature}

Marko et al. \cite{marko1995stretching} proposed an interpolation equation for force-extension experimental data of worm-like chains and has been widely used since. The force-extension equation that fits optical tweezers experiments data very well is called the MS (Marko-Siggia) equation:  
\begin{eqnarray}
    \beta f l_p = \frac{\langle X_{ee}\rangle }{L} + \frac{1}{4(1-\langle X_{ee}\rangle/L)^2} - \frac{1}{4}\;,
    \label{MS}
\end{eqnarray}
where $\beta f l_p$ is the dimensionless tension force and $\langle X_{ee}\rangle$ is the extension. MS also derives an analog of the MS equation for electric fields by equating the tension force with an effective field, and for strong stretching fields report the relationship between a field and an `effective tension' ($f_{\text{eff}}$) as, $ f_{\text{eff}} = \sigma L E/4$, where $\sigma$ is the charge density of the polymer and $l_p$ is the true persistence length.  This approximation is appropriate only in specific limits, as MS noted in the paper \cite{marko1995stretching}. For example, for finite-length chains in weak fields, the MS approximations are not guaranteed to produce the correct statistics.   This was directly addressed in the  MS paper  \cite{marko1995stretching} but is important to recognize when discussing the effective force felt by a polymer in a moderate field.

MS equation is derived under the assumption of the Weakly Bending Rod limit which implies that the transverse fluctuations are slight about the force or field stretching axis. At high fields, the leading order correction to extension is derived from this assumption, $X_{ee} \approx 1-\frac{1}{2}(\beta f l_p)^{-1/2}$. This scaling immediately implies $\beta f_{\text{eff}} = L\epsilon/4$ in the limit of high field, where $f_{\text{eff}}$ is the effective tension. In the opposite field limit, we expect that even if the field is weak the effective tension on the entire chain is not negligible. For low fields, the extension is very close to zero and grows linearly, $X_{ee}/L \approx \beta f l_p$, leading to $\beta f_{\text{eff}} = \frac{L\epsilon}{2}$. While it appears an effective tension can be defined for any value of $\epsilon$, the scaling coefficient $\beta f_{\text{eff}} =c(\epsilon)L\epsilon$ is not constant with $\epsilon$. {\color{black} The effective force as a function of the field $\epsilon$ is determined by plugging in the analytical expression for the mean end-to-end extension, $\langle X_{ee} \rangle$, obtained from the subdivided MF model equations \ref{exteqn}, into the MS equation (Equation \ref{MS}). The numerical solutions for the effective force as a function of the field are shown in Supplemental Information Figure S2} 

Figure \ref{msfec.fig} compares the FECs for the subdivided MF model with the MS model for two different values of the effective force: namely $\beta f_{\text{eff}} = L\epsilon/4$ (high fields) and $ L \epsilon/2$ (low fields). The low-field region agrees better with the effective force derived for low fields using the MS equation and similarly, the high-field FEC agrees better with the high-field effective force graph. However, none of the MS curves quantitatively captures the statistics for all three regions. Simulations agree better with the subdivided chain model in the intermediate region and although the difference may seem small in Figure \ref{msfec.fig}, due to the normalization. For long chains, the difference can be significant, especially for long reads in nanopore sequencing. The discrepancy between simulations and the MS predictions is greatest in the intermediate region. The underlying assumption of the MS model that a field can be replaced by an effective tension force is only valid in certain regions of the field and cannot be generalized with a common coefficient $c(\epsilon)$. 
 
\begin{figure}[htb!]
	\begin{center}
		\includegraphics[width=0.78\linewidth]{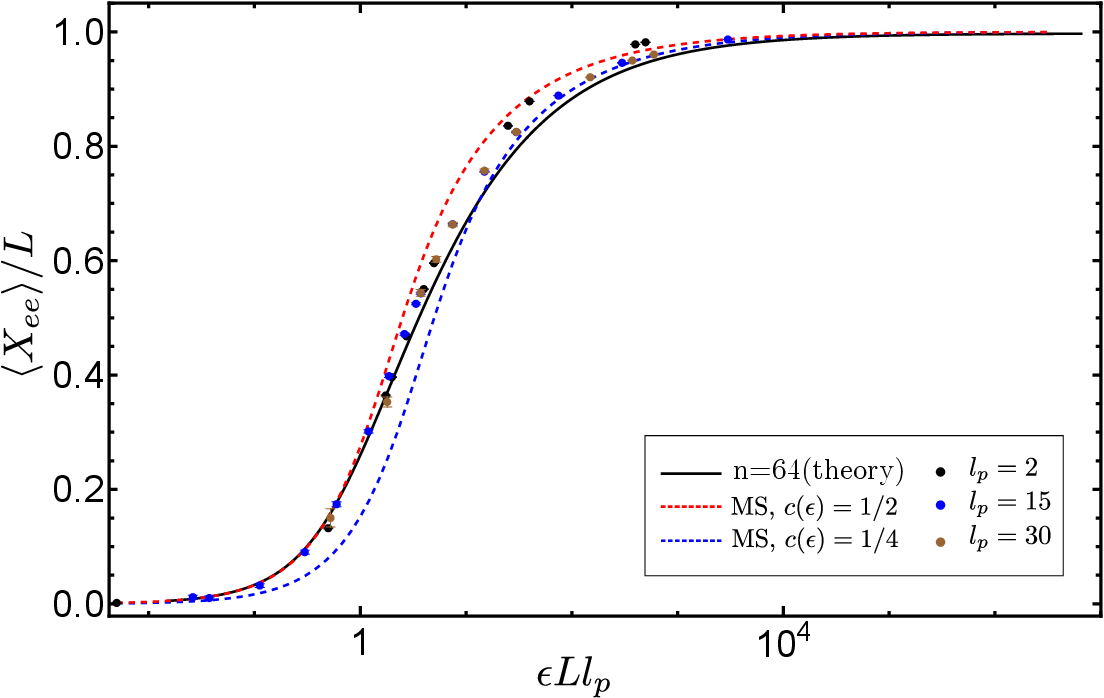}
		\caption[Comparison of the FECs for WLCs in a field with the literature]{\textbf{Comparison of the FECs for WLCs in a field with the literature}--Force-extension plots for a worm-like chain stretched under an electric field compared to the Marko-Siggia (MS) equation (Equation \ref{MS}). The solid black line is the subdivided MF model theory for $n = 64$ subdivisions. The Blue dashed line is the MS equation for the effective force, $\beta f_{\text{eff}} = \frac{L\epsilon}{4}$ and the red dashed line is the MS equation for the effective force, $\beta f_{\text{eff}} = \frac{L\epsilon}{2}$. The blue line agrees more with the subdivided MF theory for high fields, whereas the red line agrees better with low fields. Simulations are represented by solid dots with error bars for 3 different values of the stiffness parameters $l_p = 2, 15, 30$. The simulations agree better with the subdivided MF theory for the intermediate values because this model accounts for the inhomogeneity and does not approximate the field as an effective tension force. }
		\label{msfec.fig}
	\end{center}
\end{figure}

\subsection{\label{trans.sec}Transverse fluctuations}

Imaging of stiff biomolecules in an electric field has been captured experimentally \cite{maier2002elastic}, and it has been observed that the fluctuations transverse to the applied field tend to be low at the tethered endpoint and significantly greater at the free end.  For low fields, these fluctuations are larger at every point along the chain, but the fluctuations at all points along the backbone decrease as the field strength increases (as the chain becomes more elongated and aligned with the field).  To calculate the transverse fluctuations at $s = kL/n$ on the backbone we add an auxiliary field term, namely $- \alpha \int_0^{s} ds^{'} \U^{\perp}(s')$ to the Hamiltonian in Equation \ref{mfsub1}, where $\U^{\perp}(s')$ are the transverse components in $y$ and $z$ directions. It should be noted here that since $s$ is an internal point at which the transverse fluctuations are measured, the path integral is performed using the coordinates $\U$ and not the transformed one, $\V^{\perp} = \U^{\perp} - \alpha/2\lambda_i$ (see details in Supplemental Information). The Hamiltonian for calculating the transverse fluctuations after adding the auxiliary term is  
\begin{align}
	\beta H_0[\U^{\perp}_j(s)]=\delta_0 (\U_0^{\perp})^2+ \delta_L (\U_L^{\perp})^2 +\frac{l_0}{2}\int_0^L ds^{'} (\dot{\U}^{\perp})^2 &+ \sum_{j=1}^n \lambda_j\int_j ds^{'}_j (\U_j^{\perp})^2 \nonumber \\
    & - \alpha \int_0^{s} ds^{'} \U^{\perp}(s^{'})
 \label{trans1}\;.
\end{align}
The transverse fluctuations $\langle {\delta \R^{\perp}(s)}^2 \rangle = \left\langle \int_0^{s} ds^{''} \int_0^{s} ds^{'} \U^{\perp}(s^{''}) \cdot \U^{\perp}(s^{'})\right\rangle - \left\langle \int_0^{s} ds^{'}\U^{\perp}(s^{'}) \right\rangle^2$ are obtained by taking the double derivative of the free energy, $\partial^2 (-\Fc^{\perp})/\partial \alpha^2\Big{|}_{\alpha = 0}$. 

\begin{figure}[tbp!]
	\begin{center}
		\includegraphics[width=1.0\linewidth]{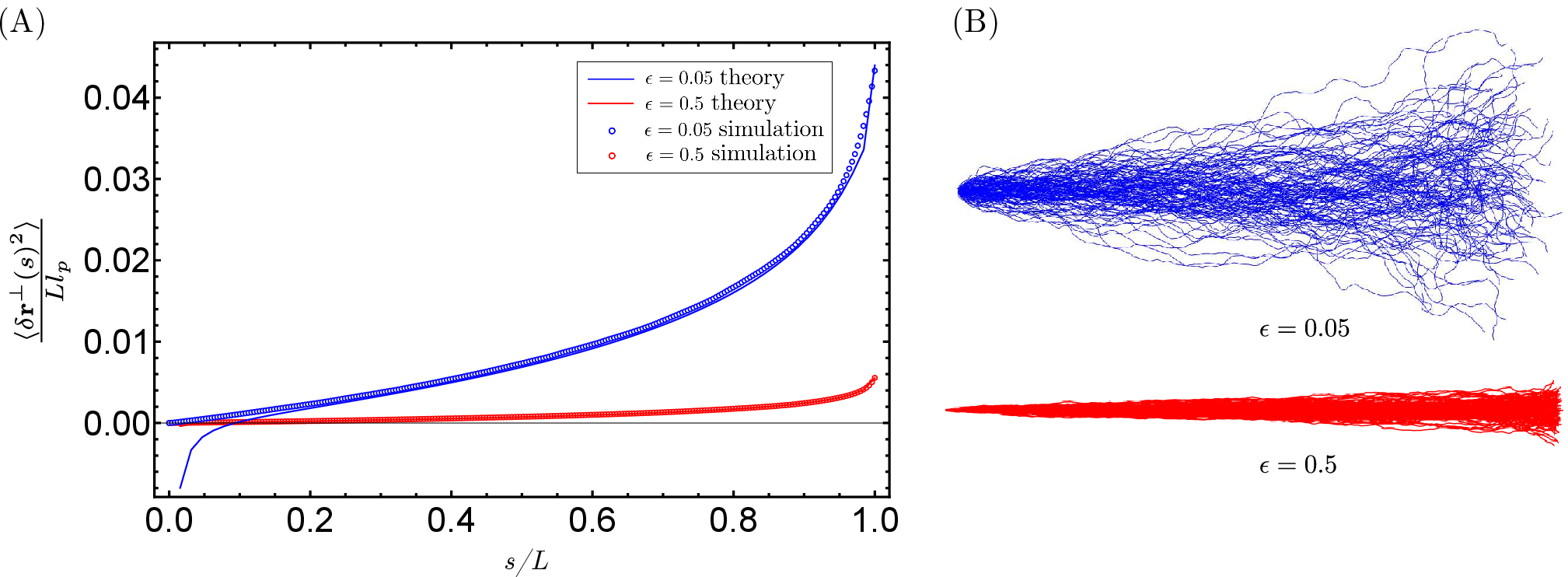}
		\caption[Transverse fluctuations in WLCs stretched using an electric field] {\textbf{Transverse fluctuations in WLCs stretched using an electric field} -- (A) Normalized transverse fluctuations, $\langle {\delta \R^{\perp}(s)}^2 \rangle/L l_p$ are plotted against the arc length ($s/L$). Solid dots represent subdivided MF theory for $n = 64$ subdivisions and lines represent simulation data. Simulations were performed for $L = 250$ monomers and $l_p = 15$ monomers. As the field increases, the transverse fluctuations decrease. For a lower value of the field (blue) the fluctuations increase along the backbone of the chain showcasing the effect of inhomogeneous stretching tension. (B) Snapshots of a trumpet (stem-flower) configurations show that the chains are aligned with the field more on the tethered end compared to the free end, which exhibits higher transverse fluctuations. The effect is more prominent for low fields (blue, $\epsilon = 0.05$).}
		\label{trans.fig}
	\end{center}
\end{figure}

The subdivided MF model can predict the transverse fluctuations in the long chain limit as shown in Figure \ref{trans.fig}. {\color{black}Figure \ref{trans.fig} shows the transverse fluctuations plotted as a function of arc length, $s$, for two different values of the applied electric field. The fluctuations are small for a high field, where the chain is fully stretched, but for a moderate value of the electric field, the fluctuations have a prominent nonlinear increasing profile with the arc length. The trumpet shape of the transverse fluctuations shown in the snapshots indicates that the chain fluctuates less on the pinned end than on the free end due to the inhomogeneous tension in the intermediate and high fields. For very low fields, the chain is hardly stretched and the transverse fluctuations are close to the fluctuations in the mean end-to-end distance squared for a worm-like chain without any force or field. The deviation near the $s = 0$ or pinned end is due to the dropping of the $\delta$ Lagrange multiplier terms for ease of calculation. }

\subsection{\label{longfluc.sec}Longitudinal fluctuations}

\begin{figure}[htb!]
	\begin{center}
		\includegraphics[width=0.8\linewidth]{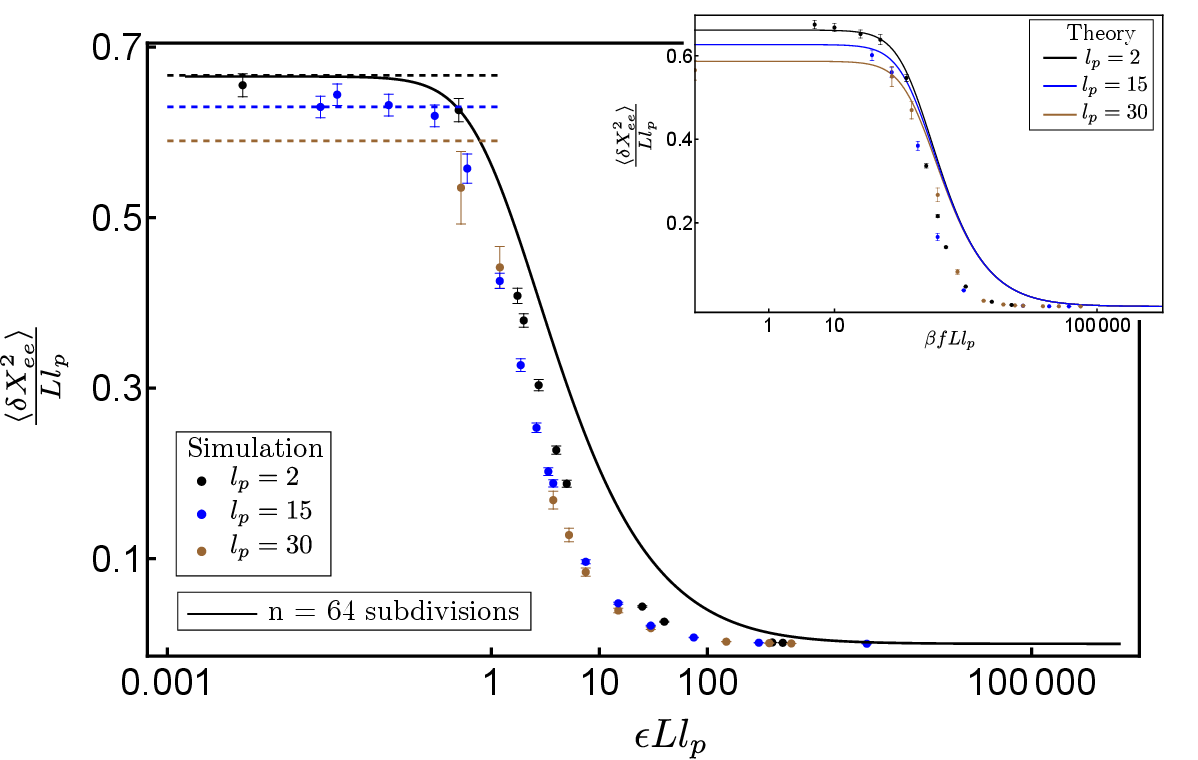}
		\caption[Longitudinal fluctuations in the mean end-to-end extension for worm-like chains stretched in an electric field] {\textbf{Longitudinal fluctuations in the mean end-to-end extension for worm-like chains stretched in an electric field} -- Subdivided MF model yields normalized longitudinal fluctuations ($\langle \delta X_{ee}^2 \rangle /Ll_p$) for a worm-like chain stretched under an electric field for $n = 64$ subdivisions (black solid line). Inset shows the longitudinal fluctuations for the case of a uniform tension $f$. Dots with error bars show simulations for three different values of stiffness parameter: $l_p = 2$ (black), $l_p = 15$ (blue),  $l_p = 30$ (brown). The mean-field model does not predict the longitudinal fluctuations correctly in the case of both the electric field and the tension (inset) due to the underlying assumption of isotropy in all directions. Finite size effects (dashed lines) are more prominent for stiff chains in low electric fields due to contributions from the exponential term in the absence of a field, with $\langle \delta X_{ee}^2 \rangle = \frac{2}{3}l_p L - \frac{2}{3}l_p^2(1-e^{-L/l_p})$. Simulations for $\kappa = 2$ and $\kappa = 15$ agree with the low electric field.}
		\label{lfluc.fig}
	\end{center}
\end{figure}

Fluctuations in the end-to-end distance also vary with the magnitude of the applied electric field. We calculate the longitudinal fluctuations for the subdivided chain model as $\langle \delta X_{ee}^2 \rangle = \frac{\partial^2 (-\Fc_n^{\alpha})}{\partial \alpha^2}\Big{|}_{\alpha = 0}$. For a WLC, the fluctuations are well known \cite{rubinstein2003polymer}, $\left\langle \int_0^L ds \int_0^L ds'u_x(s)u_x(s') \right\rangle = \frac{2}{3}l_p L - \frac{2}{3}l_p^2(1-e^{-L/l_p})$. Our subdivided model should recover these fluctuations for WLCs in the zero field limit. In the zero-field limit, the chain's fluctuations in the mean extension are also the largest because the chain is not stretched at all compared to intermediate values when the chain is partially stretched or high field values when the chain is fully stretched, and fluctuations are zero. We follow the same routine as the mean end-to-end extension in Equation \ref{exteqn} to solve for the longitudinal fluctuations in a subdivided chain. To determine the fluctuations for a subdivision indexed $j$, by solving the mean-field equations for $\{\lambda_j\}$ and the equations from the double derivative of the energy with respect to the field $\alpha$, $\lambda_j = \frac{L}{2n\langle \delta X_{ee}^2 \rangle ^j}$, where $ \langle \delta X_{ee}^2 \rangle ^j$ is the fluctuations for the subdivision indexed, $j$. The resulting longitudinal fluctuation equations for each subdivision are
\begin{eqnarray}
    -\frac{L}{n} + \frac{ \langle \delta X_{ee}^2 \rangle ^j L \epsilon^2 (1 + 3j^2 + 3n + 3n^2 - 3j(1+2n)) }{3n} + \frac{3}{2}\sqrt{\frac{\langle \delta X_{ee}^2 \rangle ^j L}{nl_0}}= 0
    \label{fluceqn}\;,
\end{eqnarray}
where $ \langle \delta X_{ee}^2 \rangle ^j$ are the transverse fluctuations. Adding the contributions from the fluctuations of all the subdivisions gives the analytical expression for the total MF longitudinal fluctuations (plotted in Figure \ref{lfluc.fig} as the solid black line). In Figure \ref{lfluc.fig}, the dashed line indicates the finite size effects from the endpoint terms. The difference between these two limits is only for the low electric field region when the field is too low to align the chain and stretch. 

In Figure \ref{lfluc.fig}, fluctuations from the subdivided chain theory are plotted for a large number of subdivisions with $n = 64$. The Monte Carlo simulations agree well with the fluctuations at low field values. For the intermediate field values, the fluctuations are over-estimated by the MF model, and for the high field values, the simulations match our model and the longitudinal fluctuations are close to zero. To understand if this discrepancy is an effect of the inhomogeneity of the tension or the MF assumption, we computed the longitudinal fluctuations of a WLC pulled with a uniform tension force $\mathbf{f}$. The discrepancy in the intermediate values of the tension force is also seen for the longitudinal fluctuations. Hinczewski et al. \cite{MikesPaper} also report the same problem: although the mean end-to-end extension is correctly predicted by the mean-field model, the underlying assumption of MF theory is that it is isotropic and does not accurately differentiate between the longitudinal and transverse fluctuations. The isotropic mean-field theory, therefore, overestimates the longitudinal fluctuations, whereas the chain fluctuates in the transverse direction more, and averaging over all three directions with equal contributions is not appropriate, ${\delta \R^{\perp}}^2 \neq \delta X_{ee}^2$, where $\R^{\perp}$ is the position components in the $y$ and $z$ directions. In the case of uniform tension, the plots for the three different values of the persistence lengths ($l_p$) are shown in the inset of Figure \ref{lfluc.fig}. 

\begin{figure}[htb!]
	\begin{center}
		\includegraphics[width=1.0\linewidth]{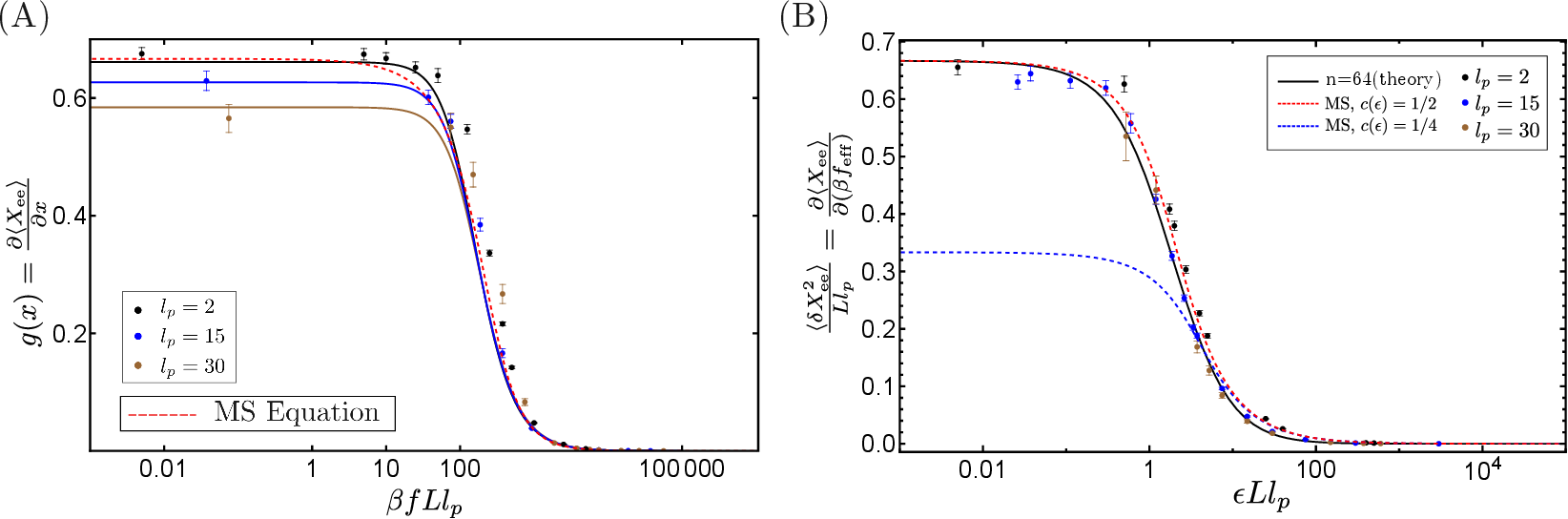}
		\caption[Comparison of the longitudinal fluctuations for WLCs in a field with the literature]{\textbf{Comparison of the longitudinal fluctuations for WLCs in a field with the literature} --Normalized longitudinal fluctuations calculated using interpolation method. (A) Interpolation equation $g(x)$ is evaluated as a derivative of the end-to-end extension with respect to the homogeneous force $x = \beta f L l_p$. The solid lines (black, blue, brown) correspond to 3 different values of the persistence length $l_p = 2,15,30$. Solid dots with error bars are simulations for $l_p = 2$ (black), $l_p = 15$ (blue), and $l_p = 30$ (brown). The predictions here agree better with the simulation than the MF fluctuations in Figure \ref{lfluc.fig}.  The finite size effects for low fields are also shown. (B) The longitudinal fluctuations for an applied field can be accurately computed by differentiating the theoretical extension (in Figure \ref{fec.fig}) with respect to an effective force, $\beta f_{eff}(\epsilon)=c(\epsilon)L\epsilon$.  For low and high fields, the effective forces reduce to $c(\epsilon)=1/2$ and 1/4 (respectively), shown in the dashed red and blue respectively.  The subdivided MF theory and the MS theory with $c(\epsilon)=1/2$ agree well with the observed fluctuations.   }
		\label{inter.fig}
	\end{center}
\end{figure}

The failure of the MF model to accurately predict longitudinal fluctuations originates from the inaccuracy of the isotropic assumption that underlies the MF model.  With no distinction between the stretching stiffness in the transverse and longitudinal directions, the theory cannot simultaneously predict the fluctuations in both directions in an analytically exact way.  In order to quantitatively predict the longitudinal fluctuations of the chain, we exploit the good agreement between simulated and predicted end-to-end distance.  To compute the fluctuations in the case of the field we use the effective force
\begin{eqnarray}
    \langle \delta X_{ee}^2 \rangle = \frac{\partial \langle X_{ee}\rangle}{\partial (\beta f_{\text{eff}})} = \frac{\partial \langle X_{ee} \rangle}{\partial \epsilon} \left( \frac{\partial (\beta f_{\text{eff}})}{\partial \epsilon} \right)^{-1}\;,
    \label{interpolationfield}
\end{eqnarray}
where $f_{\text{eff}}$ is the effective force obtained from the MS equation numerically. Equation \ref{interpolationfield} is solved numerically and plotted against the dimensionless 
electric field in Figure \ref{inter.fig} (B). 
In Figure \ref{inter.fig} (A), for a uniform tension, we find that the interpolated derivative function $ g(x) = \partial \langle X_{ee} \rangle/\partial x$, where $x = \beta f L l_p$ is the force scaled to resemble the dimensionless field $\epsilon L l_p$, agrees with the simulations better than the MF model.

 The interpolated longitudinal fluctuations have a much better agreement than the MF fluctuations for both tension and electric field because the extension is accurate with the MF model and so extracting the fluctuation as a derivative of the extension also gives a more accurate result. In Figure \ref{inter.fig} (B), the longitudinal fluctuations are also compared to the MS plots for the two values of the coefficients, namely, $f_{\text{eff}} \propto 1/4$ and $f_{\text{eff}} \propto 1/2$. The former does badly for all ranges of the electric field however the corrected effective force coefficient, $c(\epsilon) = 1/2$ is very close to the interpolated fluctuation curve.  Since the MF model predicts the FEC curves better than the fluctuations, so using the slope of the FEC curves may be the reason that the interpolated functions give more accurate statistics.

\section{\label{endelectric.sec}Discussion and conclusion}

{\color{black}In this paper, we have studied the statistics of charged worm-like chains (WLCs) stretched under uniform electric fields. We have improved on an existing mean-field theory \cite{ha1995mean,morrison2009semiflexible} used to predict the field-extension statistics for WLCs. We have shown that a homogeneous mean-field theory fails to accurately capture the statistics of the extension of the polymer. Such a homogeneous theory, with a constant stretching parameter, leads to an overstretching of the bonds and the model predicts a maximum extension of only $0.86L$ even when high electric fields are applied. We show that the overstretching and the field-extension plots can be corrected using the subdivided MF model which breaks the chain into $n$ subdivisions with different stretching stiffness $\{\lambda_j\}$. A variable stretching stiffness, imposed on the mean field level, ensures that bonds that were overstretched in the homogeneous theory have a greater resistance to stretching through the Lagrange multiplier $\lambda_j$, and that bonds that were under stretched have a lower resistance to stretching. From the field-extension curves, we also find that the effect of the inhomogeneity in tension is prominent for high applied fields. In the intermediate range, we find that the bonds closer to the tether ($s = 0$) are aligned more with the field compared to the free end ($s=L$). This is in contrast to a uniform force stretching in optical tweezers experiments, where at any point on the backbone of the chain the stretching force is the same. 

The mean-field model is an analytically tractable approach, which means a wide variety of experimental observables can also be calculated. Examples include mean end-to-end extension, longitudinal fluctuations, or transverse fluctuations. For transverse fluctuations, we recover the well-known stem-flower or trumpet statistics, arising from an unequal stretching on the end of the chain. The subdivided MF model accurately predicts the transverse fluctuations in WLCs. Nanopore sequencing devices \cite{heng2005stretching,qiu2015intrinsic, qiu2021nanopores,lu2011origins} aim to increase the precision of base sequence readings by decreasing the fluctuations while stretching DNA under intermediate electric fields. Our calculations may help us understand the molecular process of these systems better and may contribute towards improving sequencing technology. 
In the case of longitudinal fluctuations, we find that the isotropic mean-field approach fails to capture the longitudinal fluctuations both in the case of tension or field. Since the FECs are accurately predicted by the mean-field model, we propose an interpolation method that generates the fluctuation plots using the slopes of the FECs. We find that the fluctuations predicted by the interpolation method show a good agreement with the simulation data. 

Marko and Siggia's \cite{marko1995stretching} well-known force-extension equation has excellent agreement with the experimental force-extension curves for worm-like chains stretched with uniform tension forces, $\beta f$, however, fails to capture the correct statistics for fields. We find that simulations agree better with the field-extension curves obtained from the subdivided MF model in comparison to the MS equation for fields.  The MS equation is derived in the weakly bending rod limit, where the chain slightly fluctuates around the stretching axis, or in other words, higher-order transverse fluctuations are negligible. This assumption holds only for high electric fields when the chain is highly stretched. For intermediate fields, we show that the transverse fluctuations need a more exact evaluation due to the effect of the inhomogeneous tension. Longitudinal fluctuations derived from the MS equation with effective force, $f_{eff} = L\epsilon/2$ have good agreement with the simulation and are close to the interpolated fluctuations obtained from the derivative of the field-extension plots. The transverse fluctuations, however, cannot be predicted using the MS model.

In this paper, we have considered a charged polymer with a uniform charge density $\sigma$ but the model can be extended to more generalized cases of variable charges such as in block co-polymers, where the sign of the charges are alternate or there is a charge distribution, $\sigma(s)$, along the chain. In this case, the effects of charge interactions on the backbone will be an added term in the Hamiltonian and it is an interesting future direction for this project. Our model does not account for screening charges explicitly but it is an important effect for DNA in solutions, where the persistence length is modified to be the electrostatic persistence length \cite{skolnick1977electrostatic}. The persistence length in our model can be implemented as an effective length and we predict that the qualitative behavior of the statistics will not change. While applications of this work in real systems may be in nanopore sequencing devices but the effect of confinement \cite{chen2017conformational} of the nanochannels can lead to an interplay of stretching of the chain and development of hair-pin-like bends \cite{odijk2006dna,cifra2009chain} in the chain. To explore this intriguing question significant progress in simulations and theory is required but the `subdivided MF model' can serve as a starting point.}
\begin{acknowledgments}
This work was completed in part with resources provided by the Research Computing Data Core at the University of Houston. We acknowledge funding from the National Science Foundation, NSF-PHYS-2019745 supporting this work, as well as computational resources through NSF-CNS-1338099.
\end{acknowledgments}

\bibliography{electric}

\newpage

\section*{Supplemental Information}

\subsection*{S1: Calculation of free energy for a pinned, charged WLC stretched in a uniform electric field using MF model}
\label{no subdivision A matrix}

To evaluate the free energy of a charged polymer ($\sigma > 0$) in an electric field, we take a path integral approach. The Hamiltonian in Equation 2 is quadratic and can be evaluated using Gaussian integrals as follows. 
\begin{dmath*}
\Fc \propto - \log \left[ \int \Dc[\V(s)]\exp(-\beta H_{0}) \right] \nonumber\\   
= - \log \left[\int \Dc[\V(s)]\  \exp\left(-\frac{l_0}{2} \int_0^L ds \, \dot{\V}^2  -\lambda\int_0^L ds \ \V^2 \right) \exp(\mathbf{B}^T\cdot\mathbf{V} + \mathbf{C}) \nonumber \\
\exp\left(-\delta_0 \left(v_x(0) + \frac{\epsilon L}{2\lambda}\right)^2 - \delta_0 \left(v_y(0)^2 + v_z(0)^2 \right) -  \delta_L \V_L^2 \right) \right]  -  \lambda L/n - \delta_0 - \delta_L
\label{aeq: fenergy}\;.
\end{dmath*}

The quadratic terms in $\V(s)$ and $\dot{\V}(s)$ resemble the Hamiltonian of a three-dimensional classical simple harmonic oscillator (SHO) with mass, $m =  l_0$ and frequency, $\Omega = \sqrt{\frac{2\lambda}{l_0}}$. 
We can replace the quadratic part of the Hamiltonian with the simple harmonic oscillator propagator between the endpoints $\V_0$ and $\V_L$ and get
\begin{align*}
K(\V_0,\V_L,\lambda,L)  = \left(\frac{\Omega l_0}{2\pi \sinh(\Omega L)}\right)^{3/2} \exp\left(- \Omega l_0\coth(\Omega L)\frac{(\V_0^2 + \V_L^2)}{2}  + \frac{\Omega l_0}{\sinh(\Omega L)}\V_0\cdot\V_L \right).\label{aeq:SHO propagator}
\tag{S1}
\end{align*}

Substituting the propagator in \ref{aeq:SHO propagator} results in the evaluation of a well-known d-dimensional Gaussian integral 
\begin{align*}
\int \Dc [\V(s)]\  \exp(-\ \mathbf{V}^T\mathbf{A}\mathbf{V}) \exp( \mathbf{B}^T\cdot \mathbf{V} + \mathbf{C})  \propto\sqrt{\frac{\pi^d}{Det \mathbf{A}}} \exp\left( \frac{1}{4}\mathbf{B}^T \cdot \mathbf{A}^{-1} \cdot \mathbf{B}\right)\exp(\mathbf{C})\;. \label{gauformula}
\end{align*}
The transpose components of the $6 \times 1$ dimensional coordinate matrix, $\mathbf{V}^T = (v_x(0),v_x(L),\\
v_y(0),v_y(L),v_z(0),v_z(L))$. The coefficient matrix for the quadratic terms, $\mathbf{A}$, is tri-diagonal:
\[\mathbf{A}= \left(\begin{array}{cccccc}
a_{1,1} & a_{1,2} &  &   &  & 0\\
a_{2,1} & a_{2,2}& a_{2,3} &  & &\\
 & a_{3,2} &a_{3,3} & \ddots & &\\
 & &\ddots & \ddots & \ddots & a_{5,6}\\
0& && & a_{6,5}&a_{6,6}
\end{array}\right)\;,\]
where odd diagonal terms are $a_{2i+1,2i+1}  =  \delta_0 + (\Omega l_0/2) \coth(\Omega L) $ and even diagonal terms are $a_{2i,2i}  =  \delta_L + (\Omega l_0/2) \coth(\Omega L)$ for $i = 0,1,2$. The off-diagonal terms are $-\frac{\Omega l_0}{2\sinh(\Omega L)}$.

The transpose of the linear coefficient matrix $\mathbf{B}$'s is,
\begin{equation*}
    \mathbf{B}^T= \left(-\frac{l_0\epsilon}{2\lambda} - \frac{\delta_0 \epsilon L}{\lambda}\,,\;\frac{l_0\epsilon}{2\lambda}\,,\; 0\,,\;0\,,\;0\,,\;0\right)
\end{equation*}
and finally the constant term matrix is $\mathbf{C} = -\frac{\delta_0\epsilon^2L^2}{4\lambda^2} + \int_0^L ds \left(\frac{\epsilon^2 (L-s)^2}{4\lambda}-\frac{l_0\epsilon^2}{8\lambda^2}\right)$. 

\begin{figure}[htb!]
	\begin{center}
		\includegraphics[width=1.0\linewidth]{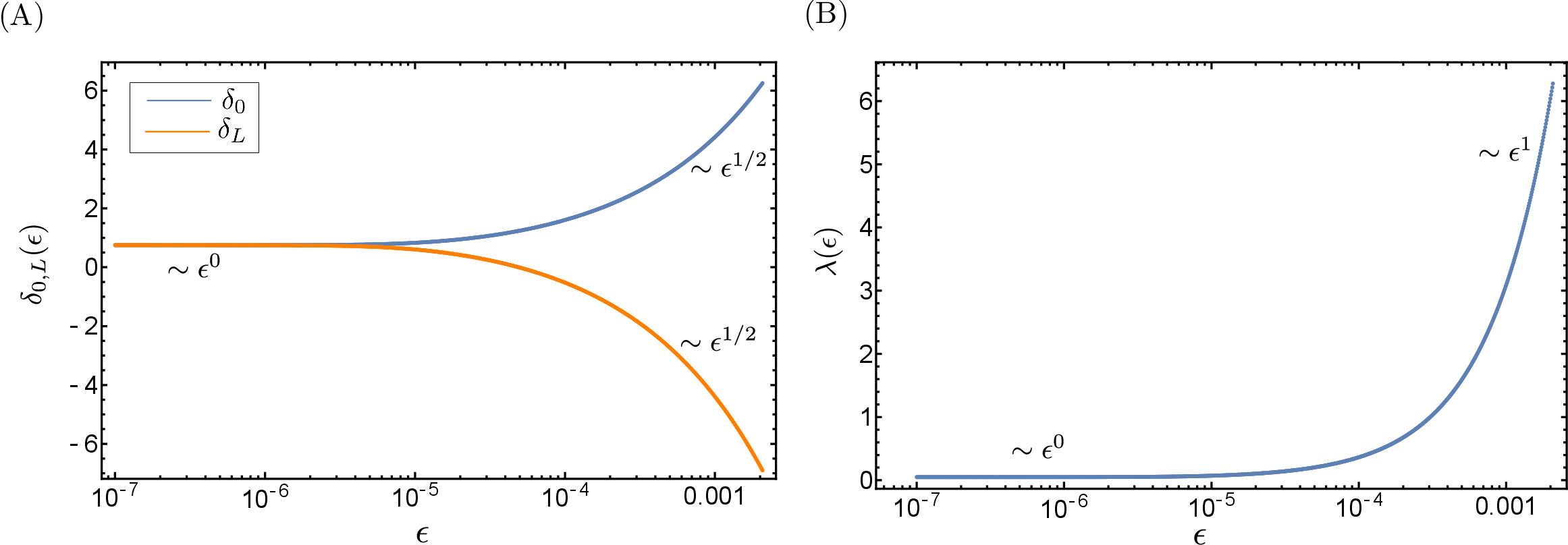}
        \captionsetup{labelformat=empty}
		\caption[Power laws of stretching parameters as a function of the electric field]{\textbf{FIG S1: Power laws of stretching parameters as a function of the electric field} --Scalings of the Lagrange multipliers $\delta_{0,L}$ and $\lambda$ as a function of the electric field $\epsilon = \sigma \beta E$. (A) The low field values for $\delta_{0,L}$ are close to the zero force or field values derived in Ha et al. \cite{ha1995mean}, $\delta = 0.75$. The asymptotic scaling is proportional to $\epsilon^{1/2}$. The difference in the sign of the $\delta$ parameters indicate the difference in the endpoint conditions. (B) Low field value for the $\lambda$ stretching parameter is close to the zero field value of $\lambda_0 = 9/8l_0 = 0.05$ for $l_0 = 3l_p/2$ and $l_p= 15$. The intermediate field region scaling for $\lambda$ is a mix of power laws between $\epsilon^0$ and $\epsilon^1$, which is the asymptotic limit.}
		\label{scaling.fig}
	\end{center}
\end{figure}

Plugging all the matrices into the d-dimensional Gaussian integral gives the partition function and the free energy. However, calculating the determinant of the tri-diagonal matrix is formidable and for long semiflexible chains, we take the assumption to the first order $\coth(\Omega L) \sim 1$ and $1/\sinh(\Omega L) \sim 0$. The assumption fails for stiff polymer chains with persistence length $l_0 >> L$. This assumption reduces $\mathbf{A}$ to a diagonal matrix whose determinant is easily calculated as the product of all the elements. The diagonal terms of the matrix are $\delta_0 + \Omega l_0/2$, or $\delta_L + \Omega l_0/2$, depending on the odd or even position. The coefficient of the propagator, $\left(\frac{\Omega l_0}{2\pi \sinh(\Omega L)}\right)^{3/2}$ expanded to the second order for the same assumption results in $\left(\frac{\Omega l_0 e^{-\Omega L}}{\pi}\right)^{3/2}$.

To obtain the scaling of the Lagrange multipliers, the mean-field equations are solved numerically. From the plots in  Figure S1, we deduce from the fits that $\lambda$ scales linearly with $\epsilon$, and the end-point Lagrange multipliers $\delta_0$ and $\delta_L$ scale as $\sqrt{\epsilon}$. Plugging these scalings back into the free energy gives us the coefficients of the scaling laws, $\lambda \sim L\epsilon/2\sqrt{3}$, $\delta_0 \sim 0.341 \sqrt{Ll_p\epsilon}$, and $\delta_L \sim -0.465 \sqrt{Ll_p\epsilon}$, where $l_p$ is the true persistence length of the chain.

\section*{S2: Calculation of free energy for a pinned, charged WLC stretched in a uniform electric field using the subdivided MF model}
\label{subdivision A matrix}
The free energy of the subdivided MF Hamiltonian in Equation 4 is calculated using Gaussian integrals. We complete squares in Equation 3 using the transformation for the $j^{th}$ subdivision, $\V_j = (v_{xj}, v_{yj}, v_{zj}) = (u_{xj} - (L-s)\epsilon/2\lambda_j, u_{yj}, u_{zj})$. The free energy is
\begin{dmath*}
\Fc_n \propto - \log \left[ \int \Dc[\V(s)]\exp(-\beta H_{0}[\V_j(s)]) \right] \nonumber\\   
= - \log \left[\int \Dc[\V(s)]\ \left\{ \sum_{j=1}^n \exp\left(-\frac{l_0}{2} \int_j ds_j \, \dot{\V_j}^2  -\lambda_j\int_j ds_j \ \V_j^2 \right) \right\} \exp(\mathbf{B}_n^T\cdot\mathbf{V}_n+ \mathbf{C}_n) \nonumber \\
\exp\left(-\delta_0 \left(v_x(0) + \frac{\epsilon L}{2\lambda_1}\right)^2 - \delta_0 \left(v_y(0)^2 + v_z(0)^2 \right) -  \delta_L \V_L^2 \right) \right]  - \sum_{j=1}^n \lambda_{j} L/n - \delta_0 - \delta_L
\label{aeq: fenergy2}
\end{dmath*},
where the sum over all the $j = 1, ...,n$ subdivisions give the total path integral of the chain, $\mathbf{B}_n$ is the coefficient matrix for the linear terms, and $\mathbf{C}_n$ is the constant term matrix, $\lambda_j$ is the stretching parameter for the $j^{th}$ subdivision, and $\lambda_1$ is the stretching parameter of the first subdivision attached to the pinned end of the chain. Just as in the undivided chain case, the simple harmonic oscillator propagator can replace the quadratic terms in the curly braces, and for each subdivision, the propagator with frequency, $\Omega_j = \sqrt{2\lambda_j/l_0}$, is given by
\begin{dmath*}
K(\V_{j-1},\V_{j},\lambda_j,L/n)  = \left(\frac{\Omega_j l_0}{2\pi \sinh(\Omega_{j} L/n)}\right)^{3/2} \exp\left(- \Omega_{j} l_0\coth(\Omega_{j} L/n)\frac{(\V_{j-1}^2 + \V_j^2)}{2}  \\
+ \frac{\Omega_{j} l_0}{\sinh(\Omega_{j} L/n)}\V_{j-1}\cdot\V_j \right).
\label{aeq:SHO propagator2}
\end{dmath*}

For a chain with $n$ subdivisions, the coefficients of the quadratic terms form the $3(n+1) \times 3(n+1)$ dimensional $\mathbf{A}_n$ matrix as follows,

\[\mathbf{A}_n= \left(\begin{array}{cccccc}
a_{1,1} & a_{1,2} &  &   &  & 0\\
a_{2,1} & a_{2,2}& a_{2,3} &  & &\\
 & a_{3,2} &a_{3,3} & \ddots & &\\
 & &\ddots & \ddots & \ddots &a_{3(n+1)-1,3(n+1)}\\
0& && &a_{3(n+1),3(n+1)-1}&a_{3(n+1), 3(n+1)}
\end{array}\right)\;,\]

and the points of subdivision spaced at every $L/n$ arc length are denoted by the $3(n+1) \times 1$ dimensional matrix, $\mathbf{V}_n$ matrix whose transpose components are,
$\mathbf{V}^T_n$$\equiv (v_{x0},v_{x1}, ..., v_{xj}, ...,\\ v_{xn},v_{y0},v_{y1},...,v_{yj},...,v_{yn},v_{z0},v_{z1},...,v_{zj},...,v_{zn})$. We take the assumption of long semiflexible chains with a large number of subdivisions and retain only diagonal terms in the $\mathbf{A}_n$ matrix. The diagonal terms corresponding to the endpoints $\V(0)$ and $\V(L)$ are, $\delta_0 + \Omega_1 l_0/2$ and $\delta_L + \Omega_n l_0/2$ respectively.  For the internal points denoting the points of subdivision, the diagonal terms are $l_0 ( \Omega_{j-1}  + \Omega_{j} )/2$. For a diagonal matrix, calculating the determinant is straightforward and it is the product of all the diagonal elements.

The linear terms form the $3(n+1) \times 1$ dimensional $\mathbf{B}_n$ matrix and the transpose components of $\mathbf{B}_n$ terms are given by,
\[\mathbf{B}_n^T= \left(\begin{array}{c}
-\frac{l_0\epsilon}{2\lambda_1} - \frac{\delta_0 \epsilon L}{\lambda_1},
\frac{l_0\epsilon}{2\lambda_1} - \frac{l_0\epsilon}{2\lambda_2},
\hdots,
\frac{l_0\epsilon}{2\lambda_{j-1}} - \frac{l_0\epsilon}{2\lambda_j},
\hdots,
\frac{l_0\epsilon}{2\lambda_n},
0,
0,
\hdots,
0
\end{array}\right)\;,\]
where only the terms in the direction of the field (here x-direction) are non-zero and all other terms are zero.

The one dimensional constant matrix, $\mathbf{C}_n = -\frac{\delta_0\epsilon^2L^2}{4\lambda_1^2} + \sum_{j=1}^n \int_j  ds_j \left(\frac{\epsilon^2 (L-s_j)^2}{4\lambda_j}-\frac{l_0\epsilon^2}{8\lambda_j^2}\right)$, where $\int_j \equiv = \intj$. 

The partition function of the subdivided MF Hamiltonian can now be evaluated using the Gaussian integral formula in Equation \ref{aeq:SHO propagator} and by replacing $\mathbf{A}, \mathbf{B}, \mathbf{C}$ with the new matrices $\mathbf{A}_n, \mathbf{B}_n, \mathbf{C}_n$. The free energy follows from the partition function,\\
$\Fc_n = -\log\left[ \int \Dc[\V(s)] \exp(-\beta H_0[\V_j(s)])\right] - \sum_{j=1}^n \lambda_{j} L/n - \delta_0 - \delta_L$. The mean-field equations $\partial \Fc_n/\partial \lambda_j = \partial \Fc_n/\partial \delta_0 =  \partial \Fc_n/\partial \delta_L = 0 $ give the distribution of the stretching parameters.

The recipe outlined in this section calculates the distribution of stretching parameters, $\lambda_j$. For a zero electric field, the stretching parameters are all equal and are that of a WLC without any stretching forces, $\lambda = 9/8l_0, \delta = 3/4$. For non-zero values of the electric field, the mean-field equations can be solved numerically. For long WLCs in the high electric field region, the scaling of the stretching parameters can be evaluated analytically. From the undivided chain scaling, $\lambda \sim L\epsilon/2\sqrt{3}$ we obtain that the stretching parameters depend on the electric field linearly, and so it is safe to assume that $\lambda_j \sim c_j \epsilon$ with coefficients, $c_j$, depending on the subdivision's location on the chain. In the high electric field limit and plugging in this scaling of $\lambda_j$'s in the free energy $\Fc_n$ we conclude that the leading order terms in the free energy are, $\Fc_n \sim \sum_{j=1}^n \int_j  ds_j \frac{\epsilon^2 (L-s_j)^2}{4\lambda_j} - \lambda_j L/n $. The mean-field equation, $\partial \Fc/\partial \lambda_j = 0$ results in, $\lambda_j = \left( \int_j ds_j (L-s_j)^2 \epsilon^2 n/4L  \right)^{1/2}$, for the $j^{th}$ subdivision's stretching parameter and so on. For a large number of subdivisions and vanishing $L/n$ length of the subdivisions, the $\lambda_j$ scaling for high electric field simplifies to $\lambda(s) \sim (L-s)\epsilon/2$ as shown in Figure 4.

\section*{\label{A3}S3: Calculation of a mean end-to-end extension using MF approach}

To calculate the ensemble average of any quantity, $\langle g \rangle$ using the definition of mean from statistical mechanics, $\langle g \rangle = \int \Dc[\V(s)] \ g \exp(-\beta H_0)/\int \Dc[\V(s)] \exp (-\beta H_0)$. The average end-to-end extension in the presence of an electric field is the total of all the bond vectors and the ensemble average for this quantity is, $\langle g \rangle = \langle \int_0^L ds u_x(s) \rangle = \left \langle \sum_{j=1}^n \int_j ds_j u_x(s_j) \right \rangle $. To make the calculation of this ensemble average simpler, we add uniform auxiliary forces \cite{ha1995mean}, $\alpha_j$'s to the Hamiltonian in Equation 3 --
\begin{align*}
	\beta H_0[\U_j(s)] &=\delta_0 \U_0^2+ \delta_L \U_L^2 +\frac{l_0}{2}\int_0^L ds \dot \U^2 + \sum_{j=1}^n \lambda_j\int_j ds_j \U_j^2 \\
    &- \boldsymbol{\epsilon} \cdot \int_0^L ds \int_0^s ds' \U(s') - \sum_{j=1}^n \alpha_j \int_j ds_j u_x(s_j) 
\tag{S2}
\label{hext1},  
\end{align*}
and complete squares as before. All the auxiliary fields are positive Lagrange multipliers and the index $j$ has been added to identify the extension relative to the $j^{th}$ subdivision. For an electric field applied in the x-direction, the new transformation in coordinates to complete the squares including the auxiliary fields is, $w_{xj} = (u_{xj} - ((L-s)\epsilon + \alpha_j)/2\lambda_j, u_{yj}, u_{zj})$. The Hamiltonian in Equation \ref{hext1} becomes quadratic with this transformation and the partition function is,  $\exp(-\Fc_{n}^{\alpha})= \exp( \log \left[ \Dc[\W(s)] \exp(-\beta H_{0}[\W_j(s)]) \right] -\sum_{j=1}^n \lambda_j L/n -\delta_0 -\delta_L)$, where $\Fc_{n}^{\alpha}$ is the free energy that is calculated using Gaussian integrals. The coefficient matrices can be derived as follows.

The quadratic coefficient matrix, $\mathbf{A}_n = \mathbf{A}_n^{\alpha}$ is unaffected by the transformation and retains all the terms for the coordinate matrix $\mathbf{W}_n^T$. The assumption of a long chain retains the diagonal terms in $\mathbf{A}_n$ as before and the product of all the elements is the determinant of the matrix. The other two matrices correspond to the linear and constant terms change and have the auxiliary fields included. For the linear terms only the endpoint terms the effect of the auxiliary fields and all other terms remain the same as $\mathbf{B}_n^T$. The endpoint terms for $w_{x} (s=0) $ and $w_{x} (s=L)$ are respectively, 
$-\frac{l_0\epsilon}{2\lambda_1} - \frac{\delta_0 (\alpha_1 + \epsilon L)}{\lambda_1}$ and $-\frac{\delta_L \alpha_n }{\lambda_n}  + \frac{l_0\epsilon}{2\lambda_n}$. The constant term matrix changes as, $\mathbf{C}_n^\alpha = -\frac{\delta_0(\epsilon L + \alpha_1)^2}{4\lambda_1^2} - \frac{\delta_L \alpha_n^2}{4\lambda_n^2}+ \sum_{j=1}^n \int_j  ds_j \left(\frac{(\epsilon(L-s_j)+\alpha_j)^2}{4\lambda_j}-\frac{l_0\epsilon^2}{8\lambda_j^2}\right)$. 

The advantage of adding auxiliary fields, $\alpha_j$ to the Hamiltonian, is we can now calculate averages easily, like the end-to-end distance, which is given by $\sum_{j=1}^{n}-\frac{\partial \Fc_n^{\alpha}}{\partial \alpha_j} \Big{|}_{\alpha_j = 0 }$. The extension for $n$ subdivisions is calculated in the main text. We should note that the distribution in the stretching parameters, $\lambda_j$, does not change because the distribution is calculated from the same mean-field equations as our subdivided MF model, $\frac{\partial \Fc_n^{\alpha}}{\partial \lambda_j} = 0\Big{|}_{\alpha_j = 0}$.

\section*{\label{A4}S4: Calculation of transverse fluctuations using the subdivided MF model}

To evaluate the transverse fluctuations from the subdivided chain Hamiltonian in Equation 8, we recognize that the calculations do not need to account for the electric field term explicitly as the perpendicular direction only feels the effect of the field through the stretching parameters. The stretching parameters have already been evaluated for the subdivided chain and for any value of the electric field the stretching parameters can be extracted and plugged into the transverse fluctuations. The transformed Hamiltonian from Equation 8 with the following transformation for arc length, $s = kL/n$ is given by $\W^{\perp}_j = \U^{\perp}_j - \alpha/2\lambda_j$ is
\begin{align*}
	\beta H_0[\W^{\perp}_j(s)]&=\delta_0 (\W_0^{\perp} + \alpha/2\lambda_1)^2 
+\delta_L (\U_L^{\perp})^2 +\frac{l_0}{2}\int_0^s ds^{'} (\dot{\W}^{\perp})^2 
+ \sum_{j=1}^k \lambda_j\int_j ds^{'}_j (\W_j^{\perp})^2 \\
&- \sum_{j = 1}^{k}  \int_0^s ds^{'} \frac{\alpha^2}{4\lambda_j} + \frac{l_0}{2}\int_s^L ds^{'} (\dot{\U}^{\perp})^2 +  \sum_{j={k+1}}^n \lambda_j\int_j ds^{'}_j (\U_j^{\perp})^2.
 \tag{S3}
\end{align*}

The harmonic oscillator propagator from this transformed Hamiltonian is used to calculate the transverse fluctuations at every point of subdivision or $ s = kL/n$, where $k = 1,2, ...,n$ is proportional to, $\left(\prod_{j=1}^{k} K(\W^{\perp}_{j-1},\W^{\perp}_{j},\lambda_j,L/n)\right) \left(\prod_{j=k+1}^{j=n} K(\U^{\perp}_{j-1},\U^{\perp}_{j},\lambda_{j},L/n)\right)$. Since the transverse fluctuations are calculated at internal points, $s$, the argument that the differences at the points of subdivision, $\eta_j = 1/\lambda_{j+1} - 1/\lambda_{j}$ become small does not give the correct statistics. The propagator expression, therefore, has to be replaced by the same coordinates for continuity, i.e. $\U$ and the propagator is then equivalent to \\
$\left(\prod_{j=1}^{k} K(\U^{\perp}_{j-1} - \alpha/2\lambda_j,\U^{\perp}_{j} - \alpha/2\lambda_j,\lambda_j,L/n)\right) \left(\prod_{j=k+1}^{j=n}K(\U^{\perp}_{j-1},\U^{\perp}_{j},\lambda_{j},L/n)\right)$. Using this propagator from the harmonic terms and along with the additional terms in the Hamiltonian Equation 8, the free energy is evaluated using Gaussian integrals as before. The transverse fluctuations is finally given by $\partial^2 (-\Fc^{\perp})/\partial \alpha^2 \Big{|}_{\alpha=0}$.

\section*{\label{A5}S5: Calculation of effective force from the MS model for WLCs stretched in electric fields}
\begin{figure}[htb!]
	\begin{center}
		\includegraphics[width=0.8\linewidth]{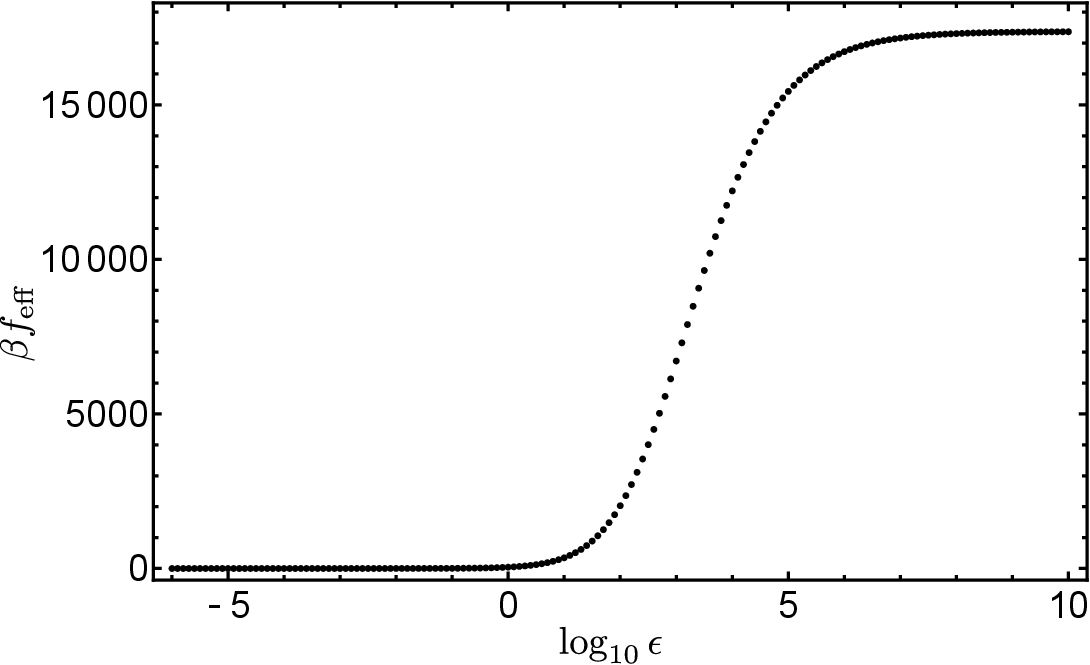}
        \captionsetup{labelformat=empty}
		\caption[Electric field as an effective tension force ]{\textbf{FIG S2: Electric field as an effective tension force} --The effective force derived from the MS equation is plotted against the field. The low and the high field values are constant but the intermediate region has a non-linear variation with the field.}
		\label{feff.fig}
	\end{center}
\end{figure}
Marko et al. \cite{marko1995stretching} proposed that the electric field can be replaced using an effective tension $f_{\text{eff}}$ from the MS equation, 
\begin{align*}
    \beta f l_p = \frac{\langle X_{ee} \rangle }{L} + \frac{1}{4(1-\langle X_{ee} \rangle)^2} - \frac{1}{4}.
    \tag{S4}
    \label{MSapp}
\end{align*}
 Plugging in the analytical expression for the total extension $\langle X_{ee} \rangle$ (obtained using the subdivided chain model equations 6) into the MS Equation \ref{MSapp} we can solve for $\beta f_{\text{eff}}$ numerically as shown in  Figure S2.

\end{document}